\documentclass[leqno]{amsart}

\usepackage[foot]{amsaddr}

\usepackage{color}
\usepackage{geometry}
\usepackage{subfiles}
\usepackage{subcaption}
\usepackage{graphicx}
\usepackage[section]{placeins}
\usepackage[linesnumbered,ruled,vlined]{algorithm2e}
\usepackage{hyperref}
\usepackage{stmaryrd}
\usepackage{amsmath,amssymb,amsfonts,amsthm,textcomp}
\usepackage{mathtools}
\usepackage{tensor}
\usepackage{multicol}
\usepackage{empheq}
\usepackage{comment}
\usepackage{tikz,pgfplots}
\usetikzlibrary{calc}
\usepackage{enumitem}
\newcommand{\lJump}{[\![}
\newcommand{\rJump}{]\!]}
\graphicspath{
  {./figures/}
  {./figures_section-1/}
  {./figures_section-2/}
  {./figures_section-3/}
  {./figures_section-adaptivity/}
  {./figures_section-examples/TimeAdap/}
  {./figures_section-examples/TimeMeshAdap/}
  {./figures_section-examples/InitialMeshAndDissip/}
}

\newfont{\tenbfsl}{cmbxti9 scaled 1200}
\newfont{\tenbbb}{msbm10}
\newfont{\svnbbb}{msbm8}

\newcommand{\bs}[1]{\boldsymbol{#1}}

\newcommand{\cl}[1]{\mathcal{#1}}

\newcommand{\fr}[2]{{\textstyle{\frac{{#1}}{{#2}}}}}

\newcommand{\bdy}{\cl{B}}

\newcommand{\tr}{\mathrm{tr}\mskip2mu}
\newcommand{\sym}{\mathrm{sym}\mskip2mu}

\newcommand{\pards}[2]{\frac{\partial{#1}}{\partial{#2}}}

\theoremstyle{remark}

\theoremstyle{definition}

\newcounter{syn}[section] 
\renewcommand{\thesyn}{\arabic{section}.\arabic{syn}}

\makeatletter
\def\threevdots{\mskip+4mu\vbox{\baselineskip2.25\p@ \lineskiplimit\z@
  \kern4.9\p@\hbox{.}\hbox{.}\hbox{.}}\mskip+3.8mu}
\makeatother

\begin{document}

\title{ A spatio-temporal adaptive phase-field fracture method}
\author{Nicolas A. Labanda$^{\diamondsuit}$}
\address{$^{\diamondsuit}$ School of Electrical
  Engineering, Computing and Mathematical Sciences, Curtin
  University, P.O. Box U1987, Perth, WA 6845, Australia}
\email{nlabanda@facet.unt.edu.ar (N.A. Labanda)}
\author{Luis Espath$^{\flat}$}
\address{$^{\flat}$ Department of Mathematics, RWTH Aachen University, Geb\"{a}ude-1953 1.OG, Pontdriesch 14-16, 161, 52062 Aachen, Germany.}
\email{espath@gmail.com (L. Espath)}
\author{Victor M. Calo$^{\diamondsuit}$}
\email{vmcalo@gmail.com (V.M. Calo)}

\date{\today}

\begin{abstract}
\noindent
We present an energy-preserving mechanic formulation for dynamic quasi-brittle fracture in an Eulerian-Lagrangian formulation, where a second-order phase-field equation controls the damage evolution. The numerical formulation adapts in space and time to bound the errors, solving the mesh-bias issues these models typically suffer.  The time-step adaptivity estimates the temporal truncation error of the partial differential equation that governs the solid equilibrium. The second-order generalized-$\alpha$ time-marching scheme evolves the dynamic system. We estimate the temporal error by extrapolating a first-order approximation of the present time-step solution using previous ones with backward difference formulas; the estimate compares the extrapolation with the time-marching solution. We use an adaptive scheme built on a residual minimization formulation in space. We estimate the spatial error by enriching the discretization with elemental bubbles; then, we localize an error indicator norm to guide the mesh refinement as the fracture propagates.  The combined space and time adaptivity allows us to use low-order linear elements in problems involving complex stress paths. We efficiently and robustly use low-order spatial discretizations while avoiding mesh bias in structured and unstructured meshes. We demonstrate the method's efficiency with numerical experiments that feature dynamic crack branching, where the capacity of the adaptive space-time scheme is apparent. The adaptive method delivers accurate and reproducible crack paths on meshes with fewer elements.


\end{abstract}

\maketitle

\tableofcontents                        


\section{Introduction}

Dynamic fracture mechanisms in brittle and quasi-brittle materials involve branching phenomena, crack arrest, crack initiation, and other non-smooth effects controlled by complex processes occurring in parallel. Pioneering experimental work~\cite{ Ravi1984a, Ravi1984b, Ravi1984c, Ravi1984d, Ramulu1985} studied dynamic fracture, stating the basic concepts of initiation, steady-state propagation, and interactions with stress waves. Building on this work, a landslide of papers proposing numerical models and formulations were published by the engineering community using cohesive elements~\cite{ Camacho1996, Pandolfi2002}, enriched discontinuous elements~\cite{ Belytschko2003}, extended finite elements~\cite{ Song2006, Song2009}, and embedded discontinuities~\cite{ Armero2009, Linder2009}, among many other approaches. Despite the advances in the field, these methods require computationally expensive features, such as interface elements; others are difficult to extend to three-dimensional simulations such as the embedded and extended approaches. In short, state-of-the-art simulation methods are incapable of circumventing the mesh-bias problem in coarse meshes.

Since the early 2000s, an alternative fracture model has been popular; it avoids strong discontinuities by introducing a phase field that models damage with its evolution coupled to the solid equilibrium of Euler-Lagrange mechanical descriptions.  This approach couples two partial differential equations that allow low-order finite elements to simulate complex crack paths efficiently when the mesh can capture the crack propagation by minimum potential paths. Francfort et al.~\cite{ Francfort1998, Bourdin2000} proposed phase-field models for fracture mechanics, which were later applied to dynamic crack propagation in several remarkable contributions~\cite{ Borden12, Hofacker13, Paul2020b, Pandolfi2021}. The main drawback of this description of dynamic fracture propagation is that it requires extremely fine meshes to capture the crack topology accurately. Many mesh adaptive algorithms seek to deliver the computational efficiency of the method by building an intelligent numerical method that avoids unnecessary refinements. The first mesh adaptive method for Euler-Lagrange formulations of steady fracture was a predictor-corrector approach~\cite{ Heister15, Areias16}. Recently,~\cite{ Paul2020} proposed an isogeometric adaptive scheme for higher-order phase-field formulations of dynamic fracture.

Irreversible fracture propagation processes require adaptive time-step strategies that control sudden energy releases and, consequently, error blow-ups. We need small-time steps to accurately reproduce the crack branching, while large steps can reproduce elastic or unloading processes. Many time-adaptive strategies for phase-field equations in the context of the Allen-Cahn and Cahn-Hilliard equations~\cite{ Gomez08, Gomez11, Liu13, Guillen14, Kastner16} exist. Most estimate the error by comparing solutions obtained with different time-accuracy integrators. Therefore, these approaches compute twice the solution of the time marching scheme, which is computationally expensive. Recently,~\cite{ Vignal17} proposed a time-adaptive method for the Cahn-Hilliard equation that estimates the truncation error of the time-marching procedure using backward difference formulas. This ingenious proposal results in a simple, cheap, and robust method for large problems, where the error estimation requires simple extrapolations from previous time-step solutions.

We believe that a unified spatial and temporal adaptivity must control the time-marching problems and avoid mesh bias. Nevertheless, only one publication deals with these challenges for Euler-Lagrange formulations for dynamic fracture problems to the best of our knowledge.  In~\cite{ Paul2020}, the authors propose a simple approach to simulate dynamic crack propagation that uses the number of Newton-Raphson iterations to control the step size; they increase the time-step size when the step converges in less than four iterations or otherwise reduce it. 

Herein, we develop a thermodynamically consistent Euler-Lagrange space-and-time adaptive formulation. We evolve the dynamical system using the generalized-$\alpha$ solving the differential systems using a staggered scheme. We estimate the formulation's temporal error from the truncation error of the second-order time-integrator using backward difference formulas. We estimate the error at the current time step by comparing it with a solution's extrapolation from the previous time steps. We build an adaptive spatial discretization using a residual minimization formulation for the phase-field equation that estimates the error by measuring the distance between low-order finite elements and a bubble enriched solution space~\cite{CIER2021114027,Cier2021,CIER2021103602,ROJAS2021113686,Marcin2020,Calo2020}. We calculate the solution error in an appropriate norm that avoids mesh bias and allows efficient refinements to reduce the total computational cost while delivering solutions insensitive to the mesh and time-step sizes.

We organize the paper as follows: Section~\ref{sec:configurationalforces} formulates the dynamics fracture problem using an energetically preserving Euler-Lagrangian approach along with a classical formulation, stating the strong and weak formulations of both cases. Section~\ref{sec:timeintegrator} details the time-integration scheme considered while section~\ref{sec:adaptive} formulates the proposed mesh-and-time-step adaptive scheme, including a detailed algorithm. These sections are the main contribution of this paper. Section~\ref{sec:numexp} presents a set of numerical experiments involving crack propagation and branching, which demonstrate the advantages of our proposal, obtaining better results in meshes with much fewer elements. We draw conclusions in Section~\ref{sec:conclusion}.

\section{Dynamic fracture modeling} \label{sec:configurationalforces}

In what follows, $\bdy$ denotes a fixed region of three-dimensional space with boundary $\partial\bdy$ oriented by an outward unit normal $\bs{n}$. We denote $\boldsymbol{H}^n(\bdy)$ as the Sobolev space of $\boldsymbol{L}^2(\bdy)$-integrable functions endowed with $n$th $\boldsymbol{L}^2(\bdy)$-integrable derivatives, $(\cdot,\cdot)_\bdy$ as the $\boldsymbol{L}^2$ inner product over the physical region $\bdy$ with boundary $\partial\bdy$, and $\langle\cdot,\cdot\rangle_{\partial\bdy}$ as the $\boldsymbol{L}^2$ inner product on the boundary $\partial\bdy$.

The phase-field theory by Fried \& Gurtin~\cite{ Fri93} augments the field equations with a momentum balance
\begin{equation}\label{eq:field.equation}
  \nabla \cdot \bs{\xi}+\pi+\gamma=0,\qquad\text{and}\qquad\nabla
  \cdot\bs{\sigma}+\bs{b}=\bs{0}, 
\end{equation}
where $\bs{\xi}$ is the microstress and $\pi$ and $\gamma$ are, respectively, the internal and external microforces, $\bs{\sigma}$ is the Cauchy stress, $\bs{b}=-\rho_{0} \ddot{\bs{u}} + \rho_{0} \bs{g}$ is the inertial and body forces, $\rho_{0}$ is the referential mass density, $\bs{u}$ is the displacement (where a superposed dot denotes time differentiation).

Restricting attention to isothermal processes, where variations in temperature $\vartheta$ are negligible, that is,
\begin{equation}
  \vartheta=\vartheta_0\equiv\mathrm{constant},
\end{equation}
the free-energy density, in terms of the internal energy $\varepsilon$ and entropy $\zeta$ densities, reads
\begin{equation}
  \psi=\varepsilon-\vartheta_0\zeta
\end{equation}
and the pointwise free-energy imbalance has the form
\begin{equation}\label{eq:pointwise.free.energy.imbalance.isothermal}
  \dot\psi+\pi\dot\varphi-\bs{\xi}\cdot\nabla\dot\varphi
  -\bs{\sigma}\colon\dot{\bs{\epsilon}}\le0.
\end{equation}

Guided by the presence of the power conjugate pairings $\pi\dot\varphi$, $\bs{\xi}\cdot\nabla\dot\varphi$, and $\bs{\sigma}\colon\dot{\bs{\epsilon}}$, with small strain tensor $\bs{\epsilon}\coloneqq\sym(\nabla\bs{u})$, we consider constitutive equations that deliver the free-energy density $\psi$, internal microforce $\pi$, microstress $\bs{\xi}$, and the Cauchy stress $\bs{\sigma}$ at each point $\bs{x}$ in $\bdy$ and time $t$ in terms of the values of the phase field $\varphi$ with its gradient $\nabla\varphi$, and its time derivative $\dot\varphi$ at that point and time.

\subsection{Thermodynamically consistent formulation}

We deal with internal constraints, using the ideas of Capriz~\cite{ Cap89} who modeled continua with microstructure. Recently, da Silva et al.~\cite{ Sil13} applied them to brittle fracture, where they additively decompose flux/stress-like quantities into a quantity constitutively determined and a powerless (orthogonal) contribution. That is,
\begin{equation}
  \pi\coloneqq\pi_{\mathrm{a}}+\pi_{\mathrm{r}},
  \qquad\bs{\xi}\coloneqq\bs{\xi}_{\mathrm{a}}+\bs{\xi}_{\mathrm{r}},
  \qquad\text{and}
  \qquad\bs{\sigma}\coloneqq\bs{\sigma}_{\mathrm{a}}+\bs{\sigma}_{\mathrm{r}}.
\end{equation}
Here, the reactive term are powerless, that is
\begin{equation}
  \pi_{\mathrm{r}}\dot{\varphi}=0,
  \qquad\bs{\xi}_{\mathrm{r}}\cdot\nabla\dot{\varphi}=0,
  \qquad\text{and}\qquad\bs{\sigma}_{\mathrm{r}}\colon\dot{\bs{\epsilon}}=0.
\end{equation}
Thus, the free-energy imbalance~\eqref{eq:pointwise.free.energy.imbalance.isothermal} specializes to
\begin{equation}\label{eq:pointwise.free.energy.imbalance.isothermal.active}
  \dot\psi+\pi_{\mathrm{a}}\dot\varphi-\bs{\xi}_{\mathrm{a}}\cdot\nabla\dot\varphi
  -\bs{\sigma}_{\mathrm{a}}\colon\dot{\bs{\epsilon}}\le0.
\end{equation}

Following Coleman \& Noll~\cite{ Col63}, we enforce the satisfaction of the dissipation inequality~\eqref{eq:pointwise.free.energy.imbalance.isothermal} in all processes. Therefore, we require that:
\begin{enumerate}[label=(\roman*),font=\itshape]
\item The free-energy density $\psi$, given by a constitutive response function $\hat{\psi}$, is independent of $\dot\varphi$:
  \begin{equation}\label{eq:free.energy.functional}
    \psi=\hat{\psi}(\varphi,\nabla\varphi,\bs{\epsilon}).
  \end{equation}
\item The active microstress $\bs{\xi}_{\mathrm{a}}$ and active Cauchy stress $\bs{\sigma}_{\mathrm{a}}$ are, respectively, given by constitutive response functions $\hat{\bs{\xi}}_{\mathrm{a}}$ and $\hat{\bs{\sigma}}_{\mathrm{a}}$ that derive from the response function $\hat{\psi}$:
  \begin{equation}\label{eq:xi.S.constitutive.relation}
    \bs{\xi}_{\mathrm{a}}=\hat{\bs{\xi}}_{\mathrm{a}}(\varphi,\nabla\varphi,\bs{\epsilon})
    =\pards{\hat{\psi}(\varphi,\nabla\varphi,\bs{\epsilon})}{(\nabla\varphi)},
    \qquad\bs{\sigma}_{\mathrm{a}}
    =\hat{\bs{\sigma}}_{\mathrm{a}}(\varphi,\nabla\varphi,\bs{\epsilon})
    =\pards{\hat{\psi}(\varphi,\nabla\varphi,\bs{\epsilon})}{\bs{\epsilon}},
  \end{equation}
  Following da Silva et al.~\cite{ Sil13}, the microstructural changes that the phase field describes are irreversible. We achieve irreversibility with the constraint $\dot{\varphi}\le{0}$, given that $\varphi=0$ represents the damaged material whereas $\varphi=1$ is undamaged. Conversely, $\nabla\dot{\varphi}$ and $\dot{\bs{\epsilon}}$ are unconstrained, thus
  \begin{equation}\label{eq:xi.S.constitutive.relation.reactive}
    \bs{\xi}_{\mathrm{r}}\coloneqq\bs{0},
    \qquad\text{and}\qquad\bs{\sigma}_{\mathrm{r}}\coloneqq\bs{0}.
  \end{equation}
  Moreover,
  \begin{equation}\label{eq:pi.constitutive.relation.reactive}
    \pi_{\mathrm{r}}\coloneqq
    \begin{cases}
      \text{arbitrary} & \text{if } \dot{\varphi} = 0,\\[4pt]
      0 & \text{if } \dot{\varphi} \le 0.
    \end{cases}
  \end{equation}
\item The internal microforce $\pi$, given by a constitutive response function $\hat{\pi}$, splits additively into a contribution derived from the response function $\hat{\psi}$ and a dissipative contribution that, in contrast to $\hat{\psi}$, $\hat{\bs{\xi}}_{\mathrm{a}}$, and $\hat{\bs{\sigma}}_{\mathrm{a}}$, depends on $\dot\varphi$ and must be consistent with a residual dissipation inequality:
  \begin{equation}\label{eq:pi.constitutive.relation}
    \left\{\,
      \begin{aligned}
        \pi=\hat{\pi}(\varphi,\nabla\varphi,\dot\varphi,\bs{\epsilon})
        &=-\pards{\hat{\psi}(\varphi,\nabla\varphi,\bs{\epsilon})}{\varphi}
        +\pi_{\textrm{dis}}(\varphi,\nabla\varphi,\dot\varphi,\bs{\epsilon}),\\[4pt]
        \pi_{\textrm{dis}}(\varphi,\nabla\varphi,\dot\varphi,\bs{\epsilon})\dot\varphi&\le0.
      \end{aligned}
    \right.
  \end{equation}
\end{enumerate}
In view of the constitutive restrictions~\eqref{eq:free.energy.functional}--\eqref{eq:pi.constitutive.relation}, the response function for the free-energy density serves as a thermodynamic potential for the microstress, the hypermicrostress, and the equilibrium contribution to the internal microforce. Therefore, a complete description of the response of a material belonging to the class in question requires scalar-valued response functions $\hat{\psi}$ and $\pi_{\textrm{dis}}$. Whereas $\hat{\psi}$ depends only on $\varphi$, $\nabla\varphi$, and $\nabla^2\varphi$, $\pi_{\textrm{dis}}$ depends also on $\dot\varphi$. Moreover, $\pi_{\textrm{dis}}$ must satisfy the residual dissipation inequality~\eqref{eq:pi.constitutive.relation}$_2$ for all possible choices of $\varphi$, $\nabla\varphi$, and $\dot\varphi$.

We now assign a suitable constitutive response for $\pi_{\mathrm{a}}$. If $\dot{\varphi}=0$, internal constraints are inactive, thus
\begin{equation}
  \pi_{\mathrm{a}}=-\pards{\hat{\psi}(\varphi,\nabla\varphi,\bs{\epsilon})}{\varphi},
  \qquad\text{if }\dot{\varphi}=0,
\end{equation}
and from~\eqref{eq:pi.constitutive.relation}, we have that
\begin{equation}
  \pi=-\pards{\hat{\psi}(\varphi,\nabla\varphi,\bs{\epsilon})}{\varphi}+\pi_{\mathrm{r}},
\end{equation}

Using~\eqref{eq:xi.S.constitutive.relation},~\eqref{eq:xi.S.constitutive.relation.reactive},~\eqref{eq:pi.constitutive.relation.reactive}, and~\eqref{eq:pi.constitutive.relation} in the field equation~\eqref{eq:field.equation}$_1$, we obtain an evolution equation
\begin{equation}\label{eq:1grade.phase.field.PDE}
  \begin{rcases}
    \text{if $\dot{\varphi}<0$, }
    &-\pi_{\textrm{dis}}(\varphi,\nabla\varphi,\dot\varphi,\bs{\epsilon})\\[4pt]
    \text{if $\dot{\varphi}=0$, }&-\pi_{\textrm{r}}
  \end{rcases}
  =\nabla
  \cdot\bigg(\pards{\hat{\psi}(\varphi,\nabla\varphi,\bs{\epsilon})}{(\nabla\varphi)}\bigg)
  -\pards{\hat{\psi}(\varphi,\nabla\varphi,\bs{\epsilon})}{\varphi}+\gamma,
\end{equation}
for the phase field. Equation~\eqref{eq:1grade.phase.field.PDE} is a nonconserved phase-field equation, a generalization of the Allen--Cahn--Ginzburg--Landau equation. Microstructural changes occur when $\dot{\varphi}<0$. Moreover, with the bracket operator $\langle x \rangle = \fr{1}{2}(x + |x|)$, the phase-field equation~\eqref{eq:1grade.phase.field.PDE} results in
\begin{equation}
  -\pi_{\textrm{dis}}(\varphi,\nabla\varphi,\dot\varphi,\bs{\epsilon})
  =\bigg\langle
  -\nabla
  \cdot\bigg(\pards{\hat{\psi}(\varphi,\nabla\varphi,\bs{\epsilon})}{(\nabla\varphi)}\bigg)
  +\pards{\hat{\psi}(\varphi,\nabla\varphi,\bs{\epsilon})}{\varphi}-\gamma\bigg\rangle,
\end{equation}
and
\begin{equation}
  -\pi_{\textrm{r}}=
  \bigg\langle\nabla
  \cdot\bigg(\pards{\hat{\psi}(\varphi,\nabla\varphi,\bs{\epsilon})}{(\nabla\varphi)}\bigg)
  -\pards{\hat{\psi}(\varphi,\nabla\varphi,\bs{\epsilon})}{\varphi}+\gamma\bigg\rangle.
\end{equation}

Here, we let the small strain tensor, with the spectral decomposion $\bs{\epsilon}\coloneqq\sum_{\iota=1}^n\epsilon^\iota\bs{m}^\iota\otimes\bs{m}^\iota$, assume the following decomposition
\begin{equation}
  \bs{\epsilon}=\bs{\epsilon}^++\bs{\epsilon}^-,
\end{equation}
where
\begin{equation}
  \bs{\epsilon}^+\coloneqq\sum_{\iota=1}^n\langle\epsilon^\iota\rangle^+\bs{m}^\iota\otimes\bs{m}^\iota,
\end{equation}
with $\bs{\epsilon}^-=\bs{\epsilon}-\bs{\epsilon}^+$. The elastic free-energy decomposes into
\begin{equation}\label{eq:elastic.energy}
  \psi_0(\bs{\epsilon})\coloneqq\psi_0^+(\bs{\epsilon})+\psi_0^-(\bs{\epsilon}),
\end{equation}
 where
\begin{equation}
  \left\{
    \begin{aligned}
      \psi_0^+(\bs{\epsilon})&=
      \dfrac{1}{2}\lambda\langle\tr\bs{\epsilon}\rangle^2+\mu\tr((\bs{\epsilon}^+)^2),\\[4pt]
      \psi_0^-(\bs{\epsilon})&=
      \dfrac{1}{2}\lambda(\bs{\epsilon}-\langle\tr\bs{\epsilon}\rangle)^2
      +\mu\tr((\bs{\epsilon}-\bs{\epsilon}^+)^2),
    \end{aligned}
  \right.
\end{equation}
where $\lambda$ and $\mu$ are the Lam\'{e} coefficients. The terms of~\eqref{eq:elastic.energy} are, respectively, the energies related to traction, compression, and shear.

In particular, we choose $\hat{\psi}$ and $\pi_{\textrm{dis}}$ according to
\begin{equation}\label{eq:free.energy.dissipation.def}
  \left\{\,
    \begin{aligned}
      \hat{\psi}(\varphi,\nabla\varphi,\bs{\epsilon})&=
      \psi_0^+(\bs{\epsilon})g(\varphi) + \psi_0^-(\bs{\epsilon}) +
      f(\varphi)
      +\fr{1}{2} g_{c} \ell^2|\nabla\varphi|^2,\\[4pt]
      \pi_{\textrm{dis}}(\varphi,\nabla\varphi,\dot\varphi,\bs{\epsilon})&=
      -\beta\dot\varphi,
    \end{aligned}
  \right.
\end{equation}
where $f$ is a function of $\varphi$, and $g_{c} >0$, $\ell>0$, $\beta>0$ are problem-specific-constants. Here, $\psi_0$ is the elastic free-energy of the undamaged material, $f$ and $g_{c} = \frac{G_c}{\ell}$ is a parameter that depends on the Griffith energy $G_c$, $\ell$ carries dimensions of length, and $\beta$ carries dimensions of (dynamic) viscosity. Moreover, $f$ and $g$ satisfy
\begin{equation}
  \forall\,0\le\varphi\le1,\quad
  \left\{
    \begin{aligned}
      f(1)=f^\prime(1)=0,&&{f}^\prime(\varphi)<0,\\[4pt]
      g(0)=0,\qquad{g}(1)=1,&&{g}^\prime(\varphi)>0,
    \end{aligned}
  \right.
\end{equation}
where $g\left( \varphi \right)$ is the degradation function. Granted that $\hat{\psi}$ and $\pi_{\textrm{dis}}$ are as defined in~\eqref{eq:free.energy.dissipation.def}, the thermodynamic restrictions~\eqref{eq:xi.S.constitutive.relation} and~\eqref{eq:pi.constitutive.relation} yield
\begin{equation}\label{eq:xi.pi.def}
  \bs{\xi}= g_{c} \ell^2\nabla\varphi,
  \qquad\pi=-\psi^{+}_{0}(\bs{\epsilon})g^\prime(\varphi)-f^\prime(\varphi)
  +
  \left\{
    \begin{aligned}
      -\beta\dot\varphi,&&\text{if $\dot{\varphi}<0$,}\\[4pt]
      \pi_{\textrm{r}},&&\text{if $\dot{\varphi}<0$.}
    \end{aligned}
  \right.
\end{equation}
with the superposed prime denoting differentiation with respect to $\varphi$. Using the particular constitutive relations~\eqref{eq:xi.pi.def} in the field equation~\eqref{eq:field.equation}, we obtain the Allen--Cahn--Ginzburg--Landau equation
\begin{equation}\label{eq:ac}
  \begin{rcases}
    \text{if $\dot{\varphi}<0$, }&\beta\dot\varphi\\[4pt]
    \text{if $\dot{\varphi}=0$, }&-\pi_{\textrm{r}}
  \end{rcases}
  =g_{c} \ell^2\triangle\varphi
  -\psi^{+}_{0}(\bs{\epsilon})g^\prime(\varphi)-f^\prime(\varphi)+\gamma,
\end{equation}
where $\triangle=\nabla \cdot\nabla$ denotes the Laplacian and $\pi_{\textrm{r}}$ specified by the right-hand-side of~\eqref{eq:ac} in case $\dot{\varphi}=0$.  In what follows, we set $\gamma=0$ and define the function $f(\varphi)$ as
\begin{equation}
  f(\varphi)\coloneqq g_{c} \dfrac{1}{2}(\varphi-1)^2.
\end{equation}

The strong form of the consistent problem of a deformable body undergoing dynamic fractures is: \textit{Find $\boldsymbol{u} \in  \mathbb{R}^{d}$ and $\varphi \in  \mathbb{R}$ such that:}
\begin{equation}\label{eq:PDE_EL.model.1}
  \left\{
    \begin{array}{r l l}
      \displaystyle \rho_{0} \ddot{\boldsymbol{u}}
      - g\left( \varphi \right) \nabla \cdot\bs{\sigma}
      &= \displaystyle \rho_{0} \boldsymbol{g}  & \text{in} \ \ \bdy \times I, \\[4pt]
      \displaystyle
      \ell^2\triangle\varphi-\dfrac{\ell \psi_0^+(\bs{\epsilon})}
      {G_c}g^\prime(\varphi)-f^\prime(\varphi) & =\begin{cases}
        \eta\dot\varphi & \text{if $\dot{\varphi}<0$}\\[4pt]
        -\pi_{\textrm{r}} & \text{if $\dot{\varphi}=0$}
      \end{cases}
                                                & \text{in} \ \ \bdy \times I, \\[4pt]
      \boldsymbol{u} \left( \boldsymbol{x} , t \right)
      & = \boldsymbol{u}_D   & \text{in}  \ \ \partial \mathcal{B}^{D} \times I,  \\[4pt]
      \bs{\sigma} \cdot \bs{n}
      & =\boldsymbol{t} & \text{in}  \ \ \partial \bdy^{N} \times I, \\[4pt]
      \nabla \varphi \cdot \boldsymbol{n}
      &= 0 & \text{in}  \ \ \partial \bdy^{N} \times I, \\[4pt]
      \boldsymbol{u} \left( \boldsymbol{x} , 0 \right)
      &= \boldsymbol{u}_0  & \text{in} \ \   \bdy,  \\[4pt]
      \dot{\boldsymbol{u}}  \left( \boldsymbol{x} , 0 \right)
      &= \dot{\boldsymbol{u}}_0  & \text{in} \ \   \bdy.
    \end{array}\right.
\end{equation}
Here, $I$ is the total time window, the traction on a Euler--Cauchy cut $\bs{t}$, where $\bs{n}$ is the outward unit normal, the parameter $\eta=\beta/g_c$,  and $\bs{g}$ is a body force. In this strong form, we assume the Cauchy stress is equal to the active stress $\bs{\sigma} = \bs{\sigma}_{a}$.

\subsection{Model reduction by a history-field variable}

The last set of equations produces a branched solution that leads to larger and more complex numerical solutions. In this sense, we introduce a history-field variable that considers a model reduction to force the irreversible nature of the process,  following~\cite{ MIEHE20102765, SARGADO2018458, Hofacker13}. Let us introduce $H$, strain energy history, given by
\begin{equation}
  H\coloneqq
  \begin{cases}
    \psi_0^+, & \text{if} \quad \psi_0^+(\bs{\epsilon}(\bs{x}))<H_{\mathrm{f}}(\bs{x}),\\[4pt]
    H_{\mathrm{f}} & \text{otherwise},
  \end{cases}
\end{equation}
where $H_{\mathrm{f}}(\bs{x}) = \max_{t\in[t = 0, t]} \psi^{+} \left( \bs{\epsilon}(\bs{x}) \right)$. Thus, the phase-field equation reduces to
\begin{equation}
  \beta\dot\varphi= g_{c} \ell^2\triangle\varphi-Hg^\prime(\varphi)-f^\prime(\varphi) .
\end{equation}

Finally, the strong form of the reduced problem of a deformable body undergoing dynamic fractures is: \textit{Find $\boldsymbol{u} \in  \mathbb{R}^{d}$ and $\varphi \in  \mathbb{R}$ such that:}
\begin{equation}\label{eq:PDE_EL.model.2}
  \left\{
    \begin{array}{r l l}
      \displaystyle \rho_{0} \ddot{\boldsymbol{u}}
      - g\left( \varphi \right) \nabla \cdot\bs{\sigma}
      &= \displaystyle \rho_{0} \boldsymbol{g}  & \text{in} \ \ \bdy \times I, \\[4pt]
      \displaystyle \eta \dot{\varphi} + \frac{\ell  H }{G_c}
      g^\prime(\varphi)
      + f^\prime(\varphi) -  \ell^2  \triangle \varphi
      &= 0  & \text{in} \ \ \bdy \times I, \\[4pt]
      \boldsymbol{u} \left( \boldsymbol{x} , t \right)
      & = \boldsymbol{u}_D   & \text{in}  \ \ \partial \mathcal{B}^{D} \times I,  \\[4pt]
      \bs{\sigma} \cdot \bs{n}
      & =\boldsymbol{t} & \text{in}  \ \ \partial \bdy^{N} \times I, \\[4pt]
      \nabla \varphi \cdot \boldsymbol{n}
      &= 0 & \text{in}  \ \ \partial \bdy^{N} \times I, \\[4pt]
      \boldsymbol{u} \left( \boldsymbol{x} , 0 \right)
      &= \boldsymbol{u}_0  & \text{in} \ \   \bdy,  \\[4pt]
      \dot{\boldsymbol{u}}  \left( \boldsymbol{x} , 0 \right)
      &= \dot{\boldsymbol{u}}_0  & \text{in} \ \   \bdy.
    \end{array}\right.
\end{equation}

We define two degradation functions $g$, the first one is a quadratic function
\begin{equation}
  g\left( \varphi \right) = \varphi^2,
\end{equation}
and the second one is a cubic function
\begin{equation}
  g\left( \varphi \right) = S (\varphi^3-\varphi^2)+3 \varphi^2-2 \varphi^3 ,
\end{equation}
where $S$ is a shape parameter that represents the sharpness of the phase-field interface.

\section{Staggered generalized-$\alpha$ time integrator} \label{sec:timeintegrator}

We evolve the partial differential equation system~\eqref{eq:PDE_EL.model.2} using a second-order generalized-$\alpha$ implicit time-marching method~\cite{ HULBERT1996175}. The weak problem statement is
\begin{equation} \label{eq:weakform}
  \left\{
    \begin{array}{l l l}
      \text{Find } \left( \boldsymbol{u},
      \varphi \right) \in \mathcal{U} \times \mathcal{P} \text{ s.t.}
      & &  \\[4pt] 
      \left( \boldsymbol{v} ; \rho_{0} \ddot{\boldsymbol{u}}  \right)_{\mathcal{B}}
      + a \left(  \boldsymbol{v}; \boldsymbol{u}, \varphi \right)
      &= \boldsymbol{f} \left(\boldsymbol{v}\right)  ;
        & \forall  \boldsymbol{v} \in \mathcal{V},  \\[4pt]
      \left(   q ; \eta \dot{\varphi}  \right)_{\mathcal{B}}
      + b \left(  q ; \boldsymbol{u}, \varphi \right)
      &= g \left(q\right) , &  \forall q \in \mathcal{Q},
    \end{array}\right.
\end{equation}
where the Lagrangian equations read
\begin{equation}
  a \left( \boldsymbol{v} ; \boldsymbol{u}, \varphi  \right)
  = \left(  \nabla  \boldsymbol{v},
    g \left( \varphi \right) \bs{\sigma}\left( \boldsymbol{u} \right)\right)_{\mathcal{B}}
  \ \ \text{and} \ \ f \left(\boldsymbol{v}\right) = \left(\boldsymbol{v}, \rho_{0}
    \boldsymbol{g} \right)_{\mathcal{B}} +  \left( \boldsymbol{v}, \boldsymbol{t}
  \right)_{\partial \mathcal{B}^{N} },
\end{equation}
while the Eulerian equations read
\begin{equation}
  b \left( q ; \boldsymbol{u}, \varphi  \right) =  \ell^{2} \left(  \nabla q ,  \nabla \varphi \right)_{\mathcal{B}} + \left( q ,  \left[ \frac{\ell  H }{G_c} \frac{d g\left(\varphi \right)}{d \varphi}+ 1 \right] \varphi \right)_{\mathcal{B}}  \ \ \text{and} \ \ g \left( q \right) = \left( q, 1 \right)_{\mathcal{B}}.
\end{equation}

The test space for the solid equilibrium equation is
\begin{equation}
  \mathcal{V} \coloneqq \boldsymbol{H}^{1}_{0}\left( \mathcal{B}\right) \coloneqq
  \left\lbrace  \boldsymbol{u} \in \boldsymbol{L}^2\left( \mathcal{B}\right)  |
    \nabla \boldsymbol{u} \in \boldsymbol{L}^2\left( \mathcal{B}\right),
    \boldsymbol{u} = \boldsymbol{u}_D \in \partial\mathcal{B}^{D} \right\rbrace .
  \label{eq:testspacelagrange}
\end{equation}
being $\boldsymbol{L}^2$ square integrables functions in the domain $\mathcal{B}$, while the test space for the phase-field equation is
\begin{equation}
  \mathcal{Q} \coloneqq \boldsymbol{H}^{1}\left( \mathcal{B}\right) \coloneqq
  \left\lbrace  \varphi \in \boldsymbol{L}^2\left( \mathcal{B}\right)
    |   \nabla \varphi \in \boldsymbol{L}^2\left( \mathcal{B}\right) \right\rbrace .
  \label{eq:testspaceuler}
\end{equation}

The integrability of equation~\eqref{eq:weakform} requires that $\ddot{\boldsymbol{u}}$ and $\boldsymbol{f}$ to belong to the dual space of the test space defined by the Riesz representation theorem $\mathcal{V}^{\ast} = \boldsymbol{H}^{-1}\left( \mathcal{B}\right)$, while $\dot{\varphi } \in \mathcal{Q}^{\ast}$. With these definitions, we introduce the following trial spaces 
\begin{equation}
  \mathcal{U} \coloneqq \left\lbrace \boldsymbol{u} \in \mathcal{V} \ \ |
    \ \ \ddot{\boldsymbol{u}} \in \mathcal{V}^{\ast} \right\rbrace  \ \ \ \
  \text{and} \ \ \ \ \mathcal{P} \coloneqq \left\lbrace \varphi  \in \mathcal{Q} \ \
    | \ \ \dot{\varphi } \in \mathcal{Q}^{\ast} \right\rbrace ,
  \label{eq:trialspace}
\end{equation}

Solving the fully coupled system~\eqref{eq:weakform} requires a considerable amount of computer resources, mainly during the space refinement. Thus, we solve a staggered scheme that solves the solid equilibrium equations independently from the phase-field equations with a Picard iteration~\cite{ Borden2012}. In every step, we first solve the equilibrium equation to obtain a displacement field; then, we solve the phase-field equation using these displacements.  This staggered technique allows us to use different time-integrators to solve each equation.

We use a semi-discrete formulation in the time interval $t_0<t_1<...<t_n<...<t_f$  and define the time step size as $\Delta t_n=t_n-t_{n-1}$. A generalized-$\alpha$ time-marching scheme for $\boldsymbol{u}\left(t_n\right)$, $\ddot{\boldsymbol{u}} \left( t_n \right)$, $\varphi  \left( t_n \right)$ and $\dot{\varphi}  \left( t_n \right)$, respectively, using $\boldsymbol{u}_n$, $\ddot{\boldsymbol{u}}_n$, $\varphi_n$ and $\dot{\varphi}_{n}$, allows us to state (for more details, see,~\cite{ chung1993time}):
%
\begin{equation} \label{eq:EqGenAlpha2}
  \begin{array}{l l }
    {\boldsymbol{u}}_{n+\alpha^{c}_f} & = {\boldsymbol{u}}_{n}  +\alpha^{c}_f \lJump {\boldsymbol{u}} \rJump , \\ [4pt]
    \ddot{\boldsymbol{u}}_{n+\alpha^{c}_m} &= \ddot{\boldsymbol{u}}_{n} + \alpha^{c}_m \lJump \ddot{\boldsymbol{u}} \rJump, 
  \end{array}
\end{equation}
for the second-order equation, while for the first-order equation we get~\cite{ JANSEN2000305}
\begin{equation} \label{eq:EqGenAlpha2bis}
  \begin{array}{l l }
    \varphi_{n+\alpha^{j}_f} &=  \varphi_{n} + \alpha^{j}_f \lJump \varphi \rJump ,\\ [4pt]
    \dot{\varphi}_{n+\alpha^{j}_m}
                             &= \dot{\varphi}_{n} + \alpha^{j}_m \lJump \dot{ \varphi } \rJump .
  \end{array}
\end{equation}
where $\lJump \bullet \rJump = \bullet_{n+1} -  \bullet_{n}$ is the variable's time increment. We expand $\boldsymbol{u}_{n+1}$ and $\varphi_{n+1}$ with Taylor series as,
\begin{align}
  \label{eq:EqGenAlpha3}
  \lJump \boldsymbol{u} \rJump
  &= \displaystyle \sum_{j = 1}^{2} \frac{{\Delta t}^{j}}{j !} \left.{\frac{\partial^j
    \boldsymbol{u}}{\partial t^j}}\right|_{n}
    +  \ \beta^{c} {\Delta t}^{2} \lJump \ddot{\boldsymbol{u}} \rJump ,
  \\ 
  \label{eq:EqGenAlpha4} \lJump \varphi \rJump
  & =  \Delta t \left( \dot{\varphi}_{n}
    + \gamma^{j} \lJump \dot{\varphi}  \rJump \right),
\end{align}
while the velocity is
\begin{align}
  \lJump \dot{\boldsymbol{u}} \rJump  &=  \Delta t \left( \ddot{\boldsymbol{u}}_{n} +  \gamma^{c} \lJump \ddot{\boldsymbol{u}}  \rJump \right) .
                                        \label{eq:taylorhyper11}
\end{align}
We define the parameters of~\eqref{eq:EqGenAlpha2}, \eqref{eq:EqGenAlpha3}, \eqref{eq:EqGenAlpha4} and~\eqref{eq:taylorhyper11} in terms of the spectral radii for the second-order equation $\rho^{c}_{\infty}$ and for the first-order equation $\rho^{j}_{\infty}$  of the amplification matrix, the only user-defined parameter, as~\cite{ Bazilevs2008}
\begin{equation}
  \boxed{
    \begin{aligned}
      \alpha^{c}_f&= \frac{1}{1+\rho^{c}_{\infty}}&&&&&&&&&
      \alpha^{j}_f&= \frac{1}{1+\rho^{j}_{\infty}} \\
      \alpha^{c}_m&= \frac{2-\rho^{c}_{\infty}}{\rho^{c}_{\infty}+1}&&&&&&&&&
      \alpha^{j}_m&= \frac{1}{2} \left( \frac{3-\rho^{j}_{\infty}}{\rho^{j}_{\infty}+1} \right)\\
      \gamma^{c} &= \frac{1}{2} + \alpha^{c}_m - \alpha^{c}_f&&&&&&&&&
      \gamma^{j} &= \frac{1}{2} + \alpha^{j}_m - \alpha^{j}_f\\
      {\beta^{c}} &= \frac{1}{4} \left( 1 + \alpha^{c}_m - \alpha^{c}_f \right)^{2}
      &&&&&&&&&&
    \end{aligned}
  }
\end{equation}

As we solve the equations in a staggered manner, we can compute the equations' spectral radii independently. In this sense, we set $\rho_{\infty} = \rho^{c}_{\infty} = \rho^{j}_{\infty}$. The staggered generalized-$\alpha$ method to solve~\eqref{eq:weakform} results in the following discrete system
\begin{equation} \label{eq:galphael}
  \left\{
    \begin{array}{r l l}
      \left(  \boldsymbol{v} ; \rho_{0} \ddot{\boldsymbol{u}}^{k+1}_{n+\alpha^{c}_m} \right)_{\mathcal{B}}
      + a \left(  \boldsymbol{v} ; \boldsymbol{u}^{k+1}_{n+\alpha^{c}_f}, \varphi^{k}_{n+1} \right)
      &= \boldsymbol{ f }_{n+\alpha^{c}_f} \left(\boldsymbol{v}\right)   ,   & \forall  \boldsymbol{v} \in \mathcal{V}\\ [4pt]
      \left( q ;  \eta \dot{\varphi}^{k+1}_{n+\alpha^{j}_m} \right)_{\mathcal{B}}
      + b \left(  q ;  \boldsymbol{u}^{k+1}_{n+1}, \varphi^{k+1}_{n+\alpha^{j}_f}  \right)
      &= g_{n+\alpha^{j}_f} \left(q\right),  & \forall  q \in \mathcal{Q} ,
    \end{array}\right.
\end{equation}
where the superscript $k$ represents the staggered iteration. Assuming the following functional linearization
\begin{align}
  a \left( \boldsymbol{v} ; \boldsymbol{u}^{k+1}_{n+\alpha^{c}_f}, \varphi^{k}_{n+1}  \right)
  &=  a \left( \boldsymbol{v} ; \boldsymbol{u}_{n}, \varphi^{k}_{n+1} \right) + \alpha^{c}_{f}  \overrightarrow{a} \left( \boldsymbol{v} ,
    \lJump {\boldsymbol{u}^{k+1}} \rJump; \boldsymbol{u}_{n} , \varphi^{k}_{n+1}  \right)  , \\
  b \left( q ;  \boldsymbol{u}^{k+1}_{n+1}, \varphi^{k+1}_{n+\alpha^{j}_f}  \right)
  &=  b \left( q ;  \boldsymbol{u}^{k+1}_{n+1}, \varphi_{n} \right)
    + \alpha^{j}_{f}  \overrightarrow{b} \left(  q , \lJump \varphi^{k+1} \rJump ;
    \boldsymbol{u}^{k+1}_{n+1} , \varphi_n \right)  ,
\end{align}
where $\overrightarrow{\bullet}$ represents the Gâteaux derivative of the functional, that is, 
\begin{equation}
  \overrightarrow{a}\left( \boldsymbol{v}, \lJump  \boldsymbol{u} \rJump ;
    \boldsymbol{u}, \varphi \right)=\dfrac{d}{d\epsilon}\left.a\left( \boldsymbol{v} ;
      \boldsymbol{u}+\epsilon \lJump  \boldsymbol{u} \rJump, \varphi\
    \right)\right|_{\epsilon=0}, 
\end{equation}
and
\begin{equation}
  \overrightarrow{b}\left( q, \lJump  \varphi \rJump ;
    \boldsymbol{u}, \varphi  \right)=\dfrac{d}{d\epsilon}\left.b\left( q ;
      \boldsymbol{u}, \varphi+\epsilon  \lJump  \varphi \rJump
    \right)\right|_{\epsilon=0}, 
\end{equation}
being $ \lJump  \boldsymbol{u} \rJump$ and $\lJump  \varphi \rJump$ proper trial functions. Substituting~\eqref{eq:EqGenAlpha2}--\eqref{eq:taylorhyper11} into equation~\eqref{eq:galphael}, the semi-discrete problem reads: \textit{Given $\boldsymbol{u}_{n}$, $\dot{\boldsymbol{u}}_{n}$, $\ddot{\boldsymbol{u}}_{n}$, ${\varphi}_{n}$, $\dot{\varphi}_{n}$ and a time-step size $\Delta t$, }
\begin{align}
  & \ \  \text{Find } \left( \lJump \boldsymbol{u} \rJump , \lJump \varphi \rJump \right) \in \mathcal{U} \times \mathcal{P} \text{ s.t.}
  & \nonumber  \\ 
  \displaystyle  & \ \  \left( \boldsymbol{v} ; \rho_{0} \lJump {\boldsymbol{u}^{k+1}} \rJump \right)_\mathcal{B}
                   + \frac{\beta^{c} \Delta t^2 \alpha^{c}_f}{\alpha^{c}_m} \overrightarrow{a} \left( \boldsymbol{v}, \lJump {\boldsymbol{u}^{k+1}} \rJump  ;  {\boldsymbol{u}_{n}}  , \varphi^{k}_{n+1} \right)
                   = \displaystyle \boldsymbol{f}^{G \alpha} \left(\boldsymbol{v}\right) ,  & \forall  \boldsymbol{v} \in \mathcal{V}, \label{eq:Galphafinal}  \\ 
  \displaystyle & \ \   \left(   q ; \eta \lJump {\varphi}^{k+1} \rJump \right)_\mathcal{B}
                  + \frac{\gamma^{j} \Delta t  \alpha^{j}_f}{\alpha^{j}_m} \overrightarrow{b} \left( q , \lJump \varphi^{k+1} \rJump ;
                  \boldsymbol{u}^{k+1}_{n+1},  \varphi_{n}   \right)  =  g^{G \alpha} \left( q\right), & \forall q \in \mathcal{Q},  \label{eq:Galphafinal2} 
\end{align}
with the following right-hand sides
\begin{align}
  \boldsymbol{ f }^{G \alpha} \left(\boldsymbol{v}\right)
  &= \frac{\beta^{c} \Delta t^2}{\alpha^{c}_m} \left[ \boldsymbol{f}_{n+\alpha^{c}_f} \left(\boldsymbol{v}\right)
    +  \rho_{0} \left(  \boldsymbol{v} ; \left(\frac{\alpha^{c}_m}{2 \beta^{c}}
    - 1 \right) \ddot{\boldsymbol{u}}_{n}
    + \frac{\alpha^{c}_m}{\beta^{c} \Delta t} \dot{\boldsymbol{u}}_{n}  \right)_\mathcal{B}
    -a \left(  \boldsymbol{v}  ; \boldsymbol{u}_{n}, \varphi^{k}_{n+1} \right) \right] ,
  \label{eq:frhs}\\
  g^{G \alpha} \left(q\right)
  &= \frac{\gamma^{j} \Delta t}{\alpha^{j}_m}\left[ g_{n+\alpha^{j}_f} \left(q\right)
    - b \left( q ; \boldsymbol{u}^{k+1}_{n+1}, \varphi_{n}  \right)
    +  \eta \left(\frac{\alpha^{j}_m}{\gamma^{j}}
    - 1 \right) \left( q ; \dot{\varphi}_{n} \right)_\mathcal{B}  \right] .
  \label{eq:grhs}
\end{align}

The algorithm~\eqref{Alg1Adap} summarizes the staggered solution of the resulting formulation, along with the space-time adaptive scheme.

\section{Fully automatic space-and-time adaptivity}
\label{sec:adaptive}

\begin{algorithm} 
  \SetAlgoLined
  \KwData{$\boldsymbol{u}_{n}$, $\boldsymbol{u}_{n-1}$ , $\boldsymbol{u}_{n-2}$, $\dot{\boldsymbol{u}}_{n}$,  $\ddot{\boldsymbol{u}}_{n}$,  $\Delta t_{n+1}$, $\Delta t_{n}$, $\Delta t_{n-1}$ and initial mesh $\mathcal{M}^{0}_{n+1}$}
  \KwResult{updated variables $\boldsymbol{u}_{n+1}$, $\dot{\boldsymbol{u}}_{n+1}$, $\ddot{\boldsymbol{u}}_{n+1}$, $\Delta t_{n+1}$ and final mesh $\mathcal{M}_{n+1}$}
  \While{$t_{n+1} \leq t_f$}{
    Initialize mesh $\mathcal{M}^{0}_{n+1}$ \;
    \While{$\| \varepsilon_{n+1} \| \geq tol_{mesh}$}{
      iteration $j \leftarrow j + 1$ \;
      Project variables of previous time-step in current mesh $  \left( \bullet \right)_{n} \left( \mathcal{M}^{j}_{n+1} \right) \leftarrow  \left( \bullet \right)_{n} \left( \mathcal{M}_{n} \right)$ \;
      \While{$ \max \left( \| \delta \boldsymbol{u} \| , \| \delta \varphi \| \right)  \geq \text{tol}_{stg} $}{
        Iteration $k \leftarrow k+1$ \;
        Calculate $\lJump \boldsymbol{u}^{k+1} \rJump$ with~\eqref{eq:Galphafinal} and $\boldsymbol{u}^{k+1}_{n+1} = \boldsymbol{u}_{n} + \lJump \boldsymbol{u}^{k+1} \rJump $   \;
        Calculate $\lJump \varphi^{k+1} \rJump$ with~\eqref{eq:Galphafinal2} and $\varphi^{k+1}_{n+1} = \varphi_{n} + \lJump \varphi^{k+1} \rJump $ \;
        Calculate $ \| \delta \boldsymbol{u} \| = \| \varphi^{k+1}_{n+1}  - \varphi^{k}_{n+1} \| $ and $ \| \delta \varphi \| = \| \varphi^{k+1}_{n+1}  -  \varphi^{k}_{n+1} \|$ \;
        \If{$ \max \left( \| \delta \boldsymbol{u} \| , \| \delta \varphi \| \right)  \leq \text{tol}_{stg} $}{
          Calculate temporal error $\boldsymbol{\tau}^{G \alpha}\left(t_{n} + \Delta t_{n+1}\right)$ with~\eqref{eq:local} \;
          Calculate $E \left(t_{n} + \Delta t_{n+1} \right)$ with~\eqref{eq:WLE} \;
          \eIf{$ E \left(t_{n} + \Delta t_{n+1} \right) \leq tol_{max}$}{
            Compute space error $\varepsilon_{n+1}$ solving~\eqref{eq:residualmin} \;
            \If{$\| \varepsilon_{n+1} \| \geq tol_{mesh}$}{
              Mark elements using $\varepsilon_{n+1}$ \;
              Refine mesh and update mesh $\mathcal{M}^{j+1}_{n+1} \leftarrow \mathcal{M}^{j}_{n+1} \left( \varepsilon_{n+1}\right)$ \;
            }
            Update current time step $t_{n+1} \leftarrow t_{n} + \Delta t_{n+1}$\;
            Update $\boldsymbol{u}_{n+1}$, $\dot{\boldsymbol{u}}_{n+1}$ and $\ddot{\boldsymbol{u}}_{n+1}$\;
            Update $\boldsymbol{u}_{n} \leftarrow \boldsymbol{u}_{n+1}$, $\dot{\boldsymbol{u}}_{n} \leftarrow \dot{\boldsymbol{u}}_{n+1}$ and $ \ddot{\boldsymbol{u}}_{n} \leftarrow \ddot{\boldsymbol{u}}_{n+1}$\;
            \If{$E \left(t_{n} + \Delta t_{n+1} \right) < tol_{min}$}{
              Increase delta step for next time increment $\Delta t_{n+1} \leftarrow  \boldsymbol{F} \left( E \left(t_{n+1}\right) ,\Delta t_{n+1}, tol \right)$}
          }{
            Reduce time-step size$\Delta t_{n+1} \leftarrow \boldsymbol{F} \left( E \left(t_{n+1}\right) ,\Delta t_{n+1}, tol \right)$\;
            Go to line $3$ and restart the staggered solution with the new time-step\;
          }
        }
      }
    }
  }
  \caption{Space-\&-time adaptivity: Staggered equilibrium \& phase-field equations' solution}
  \label{Alg1Adap}
\end{algorithm}

\subsection{Temporal adaptivity}

Since the phase-field equation is not always time-dependent, we propose an adaptive scheme in terms of the model's solid equilibrium equation, expanding the concept proposed in~\cite{ Vignal17} for second-order time derivatives. We express the local truncation error of the generalized-$\alpha$ time integrator using Taylor expansion as follows
\begin{equation}
  \boldsymbol{\tau}^{G \alpha}\left(t_{n+1}\right)
  = \frac{ \Delta t^{2}_{n+1} \left( \Delta t_{n}
      +  \Delta t_{n-1}  \right)}{6} \dddot{\boldsymbol{u}} \left(t_{n+1}\right)
  + \mathcal{O} \left( \Delta t^{4}  \right).
  \label{eq:truncation}
\end{equation}

We save the solutions $\boldsymbol{u}_{n+1}, \boldsymbol{u}_{n}, \boldsymbol{u}_{n-1}$ and $\boldsymbol{u}_{n-2}$ from the generalized-$\alpha$ scheme; then we compute the truncation error of~\eqref{eq:truncation} using the third-order backward difference formula (BDF3)
\begin{equation}
  \dddot{\boldsymbol{u}} \left(t_{n+1}\right) \approx \frac{1}{\Delta t^{2}_{n+1}}\left[ \frac{\boldsymbol{u}_{n+1}
      -\boldsymbol{u}_{n}}{\Delta t_{n+1}}
    - \left( 1 + \frac{\Delta t_{n+1}}{\Delta t_{n}} \right) \frac{\boldsymbol{u}_{n}
      -\boldsymbol{u}_{n-1}}{\Delta t_{n}}
    + \frac{\Delta t_{n+1}}{\Delta t_{n} \Delta t_{n-1}} \left( \boldsymbol{u}_{n-1}
      - \boldsymbol{u}_{n-2}\right)\right].
  \label{eq:backwarddifferenceformula}
\end{equation}

We express the local truncation error of the generalized-$\alpha$ scheme for second-order in time derivative equations by substituting~\eqref{eq:backwarddifferenceformula} into~\eqref{eq:truncation} obtaining
\begin{equation}
  \boldsymbol{\tau}^{G \alpha}\left(t_{n+1}\right) \approx \frac{ \Delta t_{n} +  \Delta t_{n-1}  }{6}\left[ \frac{\boldsymbol{u}_{n+1}-\boldsymbol{u}_{n}}{\Delta t_{n+1}} - \left( 1 + \frac{\Delta t_{n+1}}{\Delta t_{n}} \right) \frac{\boldsymbol{u}_{n}-\boldsymbol{u}_{n-1}}{\Delta t_{n}} + \frac{\Delta t_{n+1}}{\Delta t_{n} \Delta t_{n-1}} \left( \boldsymbol{u}_{n-1} - \boldsymbol{u}_{n-2}\right)\right],
  \label{eq:local}
\end{equation}
where $\Delta t_{n+1} = t_{n+1} - t_n$, $\Delta t_{n} = t_{n} - t_{n-1}$ and $\Delta t_{n-1} = t_{n-1} - t_{n-2}$ are the time increments of previous time-increments.  Finally, we compute the weighted local truncation error and use it as an error indicator~\cite{ hairer2010solving}
\begin{equation}
  E \left(t_{n+1}\right) = \sqrt{\frac{1}{N} \sum^{N}_{i=1} \left(\frac{ \boldsymbol{\tau}^{G \alpha}_{i} \left(t_{n+1}\right) }{\rho_{abs} + \rho_{rel} \max \left( \vert \boldsymbol{u}_{n+1} \vert_{i} , \vert \boldsymbol{u}_{n+1} \vert_{i} + \vert \boldsymbol{\tau}^{G \alpha} \left(t_{n+1}\right) \vert_{i} \right) }\right)^{2}},
  \label{eq:WLE}
\end{equation}
where $\rho_{abs}$ and $\rho_{rel}$ are user-defined parameters called absolute and relative tolerances, respectively. For all the examples presented in this paper,  these parameters are $\rho_{abs} = \rho_{rel} = 10^{-4}$. The time step adaptivity simply follows
\begin{equation}
  \Delta t^{k+1}_{n+1} \left(t_{n+1}\right)
  = \boldsymbol{F} \left( E \left(t_{n+1}\right) ,\Delta t^{k}_{n+1}, tol \right)
  = \rho_{tol} \left( \frac{tol}{E \left(t_{n+1}\right)}\right)^{\frac{1}{2}} \Delta t^{k}_{n+1} ,
  \label{eq:DT}
\end{equation}
where $k$ is the time-step refinement level and $\rho_{tol} $ is a safety factor parameter that we set to $0.9$ for all applications. We summarize the time-adaptive scheme as a global procedure in Algorithm~\ref{Alg1Adap}, where we define two tolerances $tol_{max}$ and $tol_{min}$ that limit the range of reduction or increments of the time-step size.

\subsection{Spatial adaptivity}

We propose an adaptive residual minimization method, which measures the error in a proper dual norm.  We use low-order elements to approximate the solution and enrich the solution space to estimate the solution error with bubbles. These bubbles allow us to localize the error measurement and guide refinement. The phase-field equation and its characteristic length bound the mesh size; thus, our procedure seeks to avoid mesh bias as the crack propagates by estimating the phase-field error. The phase-field is a scalar function; thus, the spatial error control uses the smallest representative equation system to estimate the error delivering significant computational savings.

First, we introduce a conforming partition $\mathcal{M}_{h} = \left\lbrace \mathcal{T}_{i} \right\rbrace^{N}_{i=1} $ of the domain $\mathcal{B}$ into $N$ disjoint finite elements $ \mathcal{T}_{i}$, such that
\begin{align}
  \mathcal{B}_{h} = \cup^{N}_{i=1} \mathcal{T}_{i} \qquad \text{satisfies} \qquad
  \mathcal{B} = \text{int}\left( \mathcal{B}_{h} \right).
\end{align}

We define $\mathbb{P}^{p}\left(\mathcal{T}_{i} \right)$, with $p \geq 0$, the set of polynomials of degree $p$ defined on the element $\mathcal{T}_{i}$, and consider the following polynomial space 
\begin{equation}
  \mathbb{P}^{p}\left(\mathcal{B}_{h} \right) = \left\lbrace \varphi \in
    \boldsymbol{L}_{2} \left( \mathcal{B} \right)  \ \ \vert \ \
    \varphi\vert_{\mathcal{T}_{i}} \in \mathbb{P}^{p}\left(\mathcal{T}_{i} \right),
    \forall i=1,...,N \right\rbrace ,
\end{equation}

Let ${\mathcal{Q}_{B}}_{h}$ be a finite-dimensional space built with functions in $\mathcal{B}_{h}$, the residual minimization problem for the phase-field equation considering the generalized-$\alpha$ time integrator reads
\begin{equation}
  \left\{
    \begin{array}{l l l}
      \text{Find $\left( \varepsilon_{h}, \varphi_{h} \right) \in
      {\mathcal{P}_{B}}_{h} \times \mathcal{P}_{h}$ such that} & & \\ [4pt]
      \left(  w_{h} ; {\varepsilon}_{h} \right)_{\mathcal{Q}^{\Delta t}_h}
      + b^{G \alpha}_{h} \left( w_{h}  ; {\boldsymbol{u}_{h}}_{n+1},  \varphi_{h}    \right)
                                                               &= g^{G \alpha}_{h}\left( w_{h}   \right)
                                                                 &, \forall w_{h} \in   {\mathcal{Q}_{B}}_{h} \\ [4pt]
      b^{G \alpha}_{h} \left( q_{h}  ;  {\boldsymbol{u}_{h}}_{n+1}, {\varepsilon}_{h}  \right)
                                                               &= 0  & ,   \forall q_{h} \in   \mathcal{Q}_{h},
    \end{array}\right.
  \label{eq:residualmin}
\end{equation}
where  ${\mathcal{P}_{B}}_{h}$ is the discrete bubble enriched space
\begin{equation}
{\mathcal{P}_{B}}_{h} =   \left\lbrace \varphi_{h}  \in {\mathcal{Q}_{B}}_{h} \ \ | \ \ \dot{\varphi }_{h} \in  {\mathcal{Q}_{B}}^{*}_{h} \right\rbrace,
  \label{eq:enrichedspacebuuble}
\end{equation}
and $\left( \cdot,\cdot  \right)_{\mathcal{Q}^{\Delta t}_h}$ is the inner product induced by the norm
\begin{equation}
  \| \varphi_{h} \|^{2}_{\mathcal{Q}^{\Delta t}_h} = \eta \| \varphi_{h}
  \|^{2}_{\boldsymbol{L}_2} + \frac{\alpha^{j}_f \Delta t \gamma^{j}}{\alpha^{j}_m} \| \varphi_{h} \|^{2}_{\mathcal{Q}_h},
  \qquad \text{and} \qquad \| \varphi_{h} \|^{2}_{\mathcal{Q}_h}
  = \left( 1 + \frac{l_{0}}{G_{c}} \right) \| \varphi_{h} \|^{2}_{\boldsymbol{L}_2}
  + l^{2}_{0}  \| \nabla \varphi_{h} \|^{2}_{\boldsymbol{L}_2} ,
  \label{eq:norm}
\end{equation}
being the functional 
\begin{equation}
  b^{G \alpha}_{h} \left(  w_{h}  ;   {\boldsymbol{u}_{h}}_{n+1},  \varphi_{h}   \right)
  = \left(  q_{h}  ;  \eta \lJump {\varphi}_{h} \rJump  \right)_\mathcal{B}
  + \frac{\gamma^{j} \Delta t  \alpha^{j}_f}{\alpha^{j}_m} \overrightarrow{b_{h}} \left( q_{h} ,   \lJump \varphi_{h} \rJump  ;  {\boldsymbol{u}_{h}}_{n+1}, {\varphi_{h}}_{n}   \right),
\end{equation}
is the left-hand side~\eqref{eq:Galphafinal2} and $\varepsilon_{h} = R^{-1}_{{\mathcal{Q}_{B}}_{h}} \left( g^{G \alpha}_{h}\left(w_{h}\right) - b^{G \alpha}_{h} \left(  w_{h} ;  {\boldsymbol{u}_{h}}_{n+1}, \varphi_{h}  \right) \right) \in {\mathcal{Q}_{B}}_{h} $ is an  error representation. The residual representation also has the following property
\begin{equation}
  \| {\boldsymbol{\varepsilon}_{h}}_{n+1} \|_{\mathcal{Q}^{\Delta t}_h}
  = \| B^{G \alpha}_{h} \left( {\theta_h}_{n+1}
    - {\varphi_h}_{n+1} \right) \|_{{\mathcal{Q}^{\Delta t}_h}^{*}},
\end{equation}
where ${\theta_h}_{n+1} \in {\mathcal{P}_{B}}_{h}$ is the solution of the phase-field equation~\eqref{eq:Galphafinal2} in the bubble enriched space, while $ B^{G \alpha}_{h} :  {\mathcal{Q}_{B}}_{h,\#} \rightarrow  {\mathcal{Q}_{B}}^{*}_{h}$ is defined as ~\cite{ Calo2020}:
\begin{equation}
  \left\langle w_{h} ; B^{G \alpha}_{h}  {z}  \right\rangle_{{\mathcal{Q}_{B}}_{h} \times {\mathcal{Q}_{B}}^{*}_{h}}
  = b^{G \alpha}_{h} \left(  w_{h}  ; \boldsymbol{u}_{h} ,  {z}    \right).
\end{equation}

\begin{figure} [h]
  \begin{subfigure}[b]{0.325\linewidth}
    \centering%
    {\includegraphics[scale=1.2]{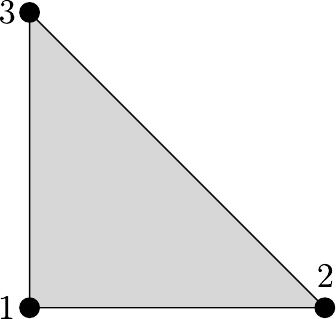}}
    \caption{Space $\mathcal{Q}$.}
  \end{subfigure}
  \begin{subfigure}[b]{0.325\linewidth}
    \centering%
    {\includegraphics[scale=1.2]{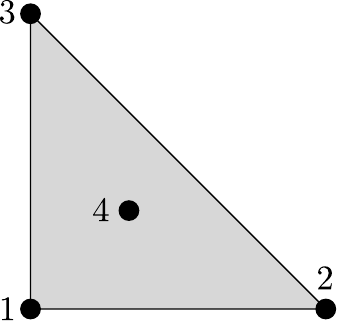}}
    \caption{Space $\mathcal{Q}_{B}$.}
  \end{subfigure}
  \begin{subfigure}[b]{0.325\linewidth}
    {\includegraphics[scale=0.6]{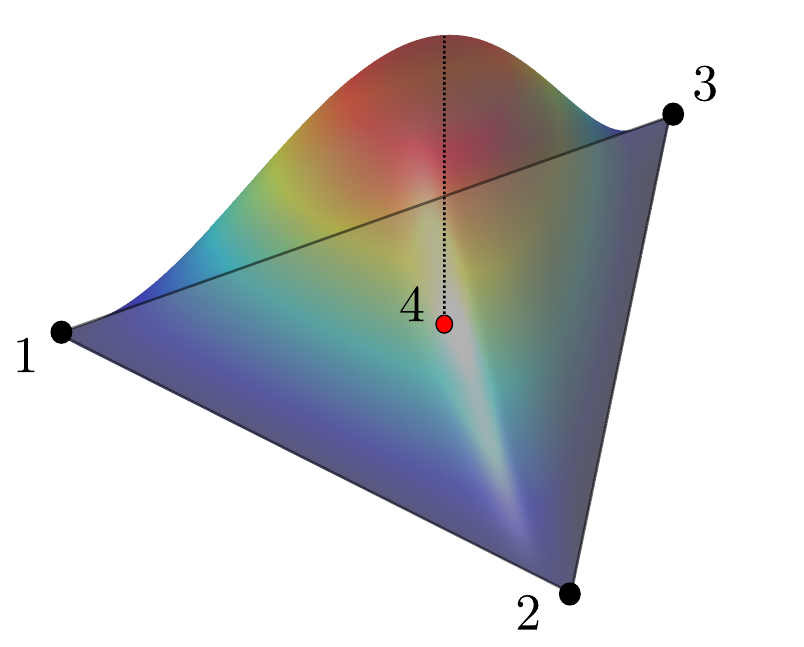}}
    \caption{Bubble shape function.}
  \end{subfigure}
  \caption{Low order space $\mathcal{Q}$ and bubble enriched space $\mathcal{Q}_{B}$ for 2D.}
  \label{Fig:2Dtech}
\end{figure}

The bubble enriched space ${\mathcal{P}_{B}}_{h} \subset {\mathcal{Q}_{B}}_{h}$ allows us to localize the error estimation to mark and refine the mesh in terms of the phase-field error. Figure~\ref{Fig:2Dtech}~(a) shows a linear triangular element while Figures~\ref{Fig:2Dtech}~(b) and~(c) show a third-order bubble function that enriches the discrete space.

In all examples, we approximate all unknown fields with linear functions  and then enrich these spaces with bubbles to estimate the spatial error.


\subsection{Adaptive space-and-time procedure}

This section describes the adaptive procedure to provide the reader with all the necessary tools for a straightforward implementation. We implement the solver we use in this paper in the open-source package FEniCS project~\cite{ alnaes2015fenics} combined with the FIAT package to deal with different quadratures~\cite{ Kirby2004}. The Algorithm~\ref{Alg1Adap} describes a general calculation layout for the proposed model.

We obtain a reliable user-independent numerical method for simulating dynamic fracture processes. We assume the user provides the algorithm with an initial mesh $\mathcal{M}^{0}_{n+1}$  as an iteration starting point. Also, the user needs to define an initial time-step size $\Delta  t^{0}$ to start the calculation procedure. Thereon,  the time-adaptive approach bounds the error temporal error between $tol_{min}$ and $tol_{max}$.  Furthermore, the user needs to specify two more tolerances, $tol_{stg}$ that controls the staggered solution of Equations~\eqref{eq:Galphafinal} and~\eqref{eq:Galphafinal2}, and $tol_{mesh}$ that control the spatial mesh refinement process.

In all numerical experiments, we select for refinement those elements within a certain percentage of the element with maximum error. We summarize the approach as follows
\begin{equation}
  \text{refine all } \mathcal{T}_{i} \in \mathcal{M}_{h} \text{ such that }
  \varepsilon_{n+1}\left(\mathcal{T}_{i}\right) \geq \chi \max \left( \varepsilon_{n+1} \right),
\end{equation}
where $\chi \in \left[0,1\right]$ is a user-defined parameter. When $\chi$ approaches $0$, we refine fewer elements in each iteration, while as it tends to $1$, we refine most elements. Our experience shows that this element-selection method delivers the most efficient discretizations  when for $\chi \in \left[ 0.1 , 0.3 \right]$.

\section{Numerical Experiments} \label{sec:numexp}


\subsection{Significance of an error-based time-adaptive scheme}

In this section, we study the dynamic branching problem, focusing on the impact of the time-adaptive scheme to avoid spurious solutions. We use.a fixed, fine mesh to isolate the influence of the time adaptive scheme on the process.

\begin{figure} [h]
\centering%
{\includegraphics[scale=0.5]{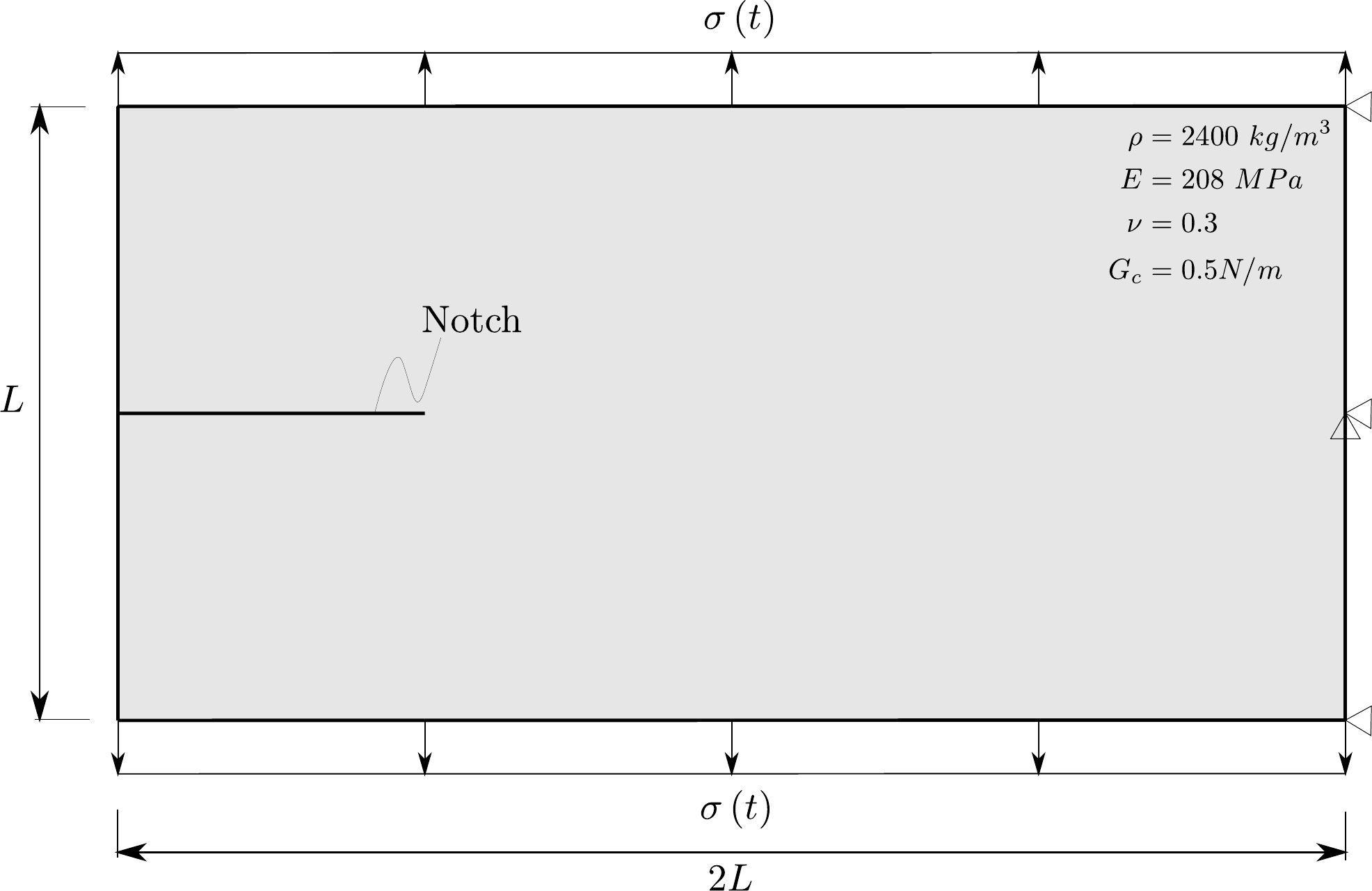}}
\caption{Dynamic fracture in a notched plate. Problem setup.}
\label{Fig:BranchingSetup}
\end{figure}

Figure~\ref{Fig:BranchingSetup} shows a notched plate of dimension $1\, m \ \text{x} \ 2\, m$ that contains an initial planar notch of $50\,cm$. We subject the plate to a vertical traction force $\sigma \left(t\right) = 10KN$. A constant uniform mesh of $262,144$ elements with $263,938$ degrees of freedom for the displacement field and $131,969$ for the phase-field, totaling $395,907$ for the overall problem. Figure~\ref{Fig:BranchingSetup} also includes the material parameters, being the elastic modulus $E = 208 \,MPa$, Poisson modulus $\nu = 0.3$, fracture energy $G_c = 0.5 \,N/m^2$ and density $\rho = 2,400\, kg/m^3$. We set the characteristic length equal to the minimum element size $\ell = 5mm$ and use a quadratic degradation function.  We denote the minimum element size the mesh size.

\begin{figure}[h!]
\centering%
\begin{subfigure}[b]{0.425\linewidth}
\centering
{\includegraphics[height=4.4cm,trim={0cm 0cm0 0cm
    0cm},clip]{CoarseBranchSimWO.pdf}}
\vspace{.2cm}
\caption{Phase field, $t = 25 \, ms$}
\end{subfigure}
\begin{subfigure}[b]{0.565\linewidth}
\centering%
{\includegraphics[height=4.5cm,trim={0 0 0 0},clip]{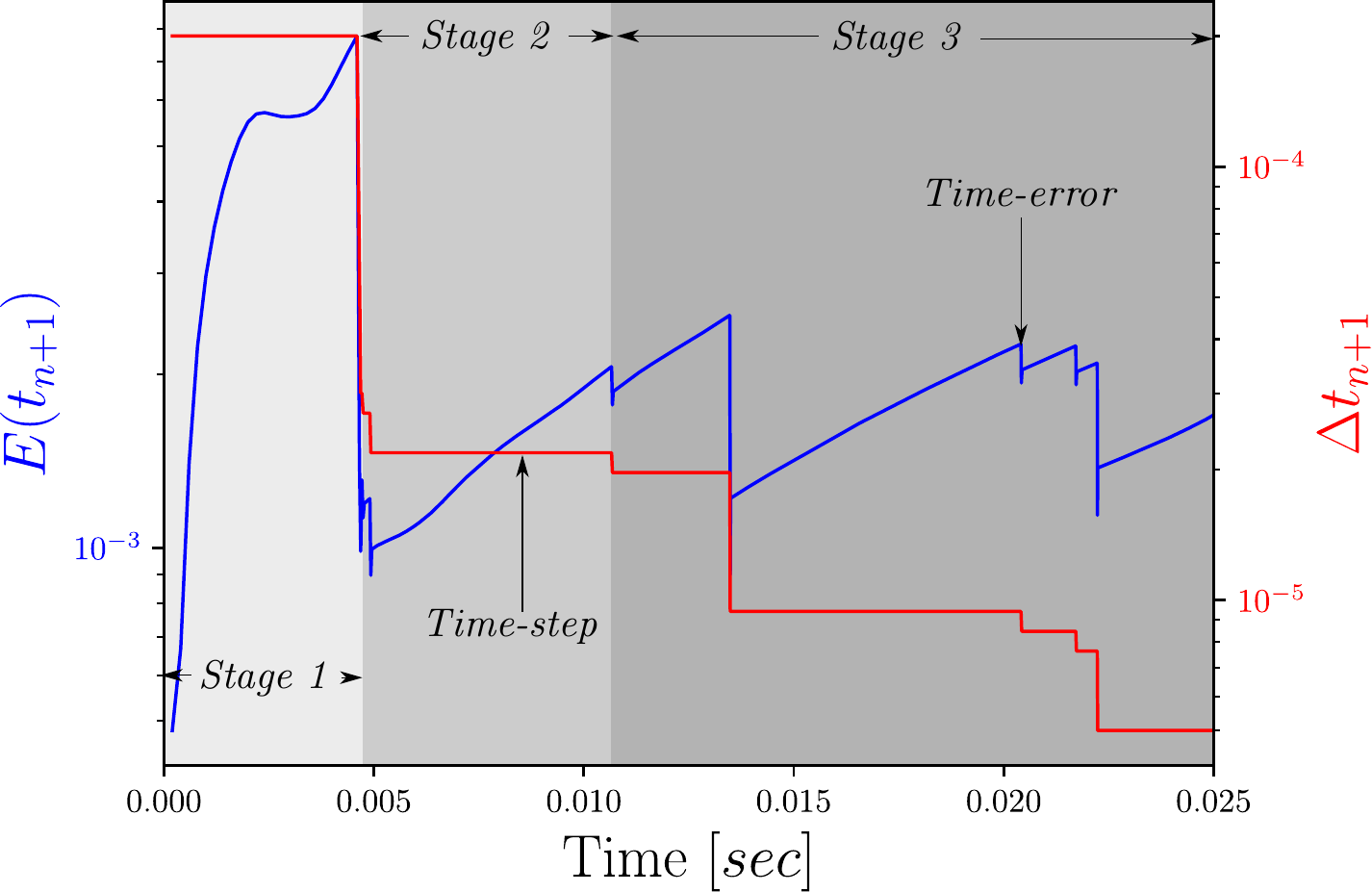}}
\caption{Error $E \left( t_{n+1} \right)$ \& time-step size $\Delta t$ evolution}
\end{subfigure}
\caption{Dynamic fracture branching: time adaptivity based on iteration count}
\label{Fig:PhaseFieldTimeAdapWO}
\end{figure}

\begin{figure}[h!]
  \centering%
  \begin{subfigure}[b]{0.32\linewidth}
    \centering%
    {\includegraphics[height=2.72cm,trim={0cm 0cm0 0.0cm 0cm},clip]{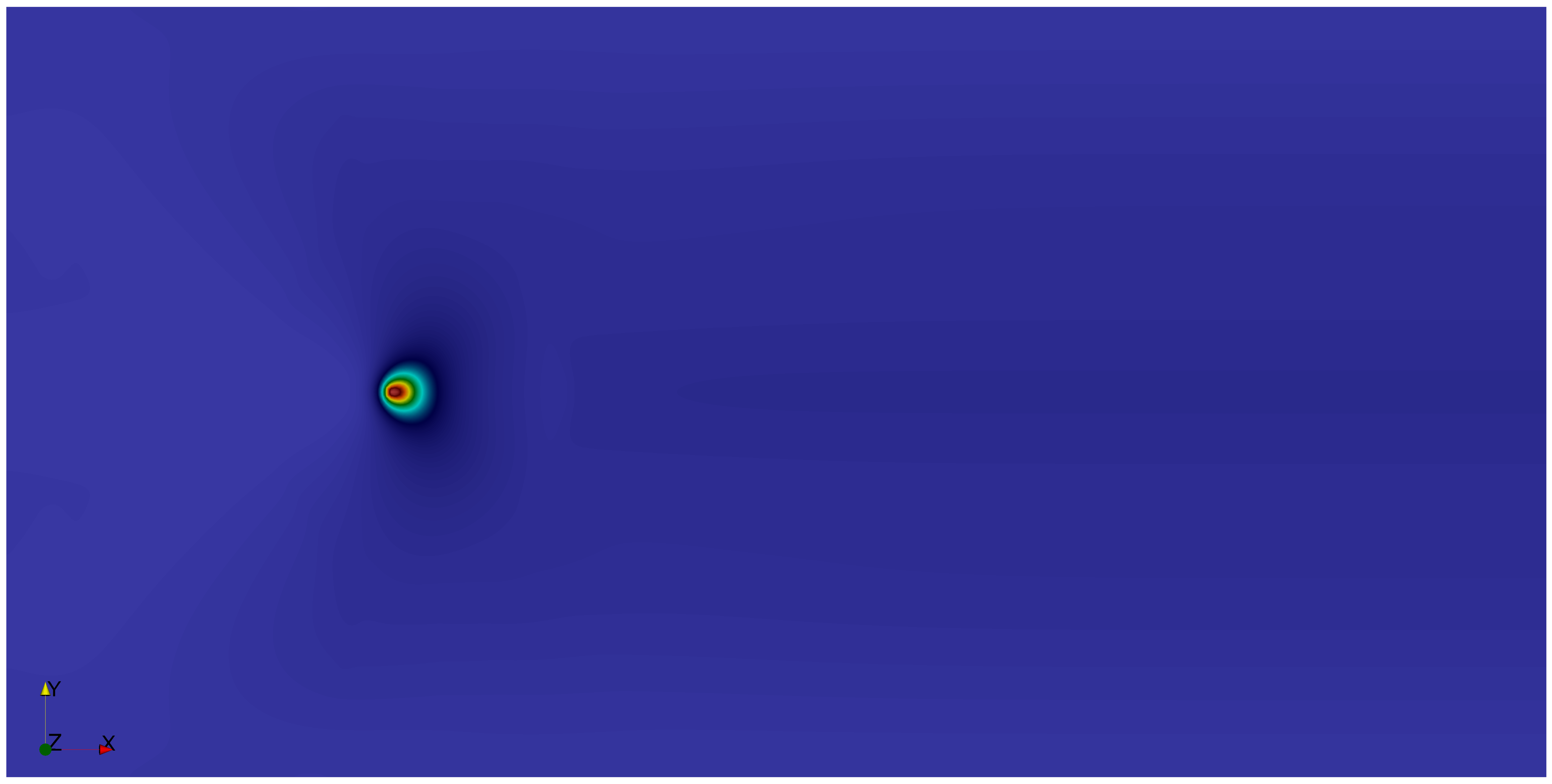}}
    \caption{Phase field, $t = 5 \, ms$}
  \end{subfigure}
  \begin{subfigure}[b]{0.32\linewidth}
    \centering%
    {\includegraphics[height=3.05cm,trim={0cm 0cm  0cm 0cm},clip]{BranchSimt001.pdf}}
    \caption{Phase field, $t = 10 \,ms$}
  \end{subfigure}
  \begin{subfigure}[b]{0.32\linewidth}
    \centering%
    {\includegraphics[height=2.75cm,trim={0cm 0cm0 0cm 0cm},clip]{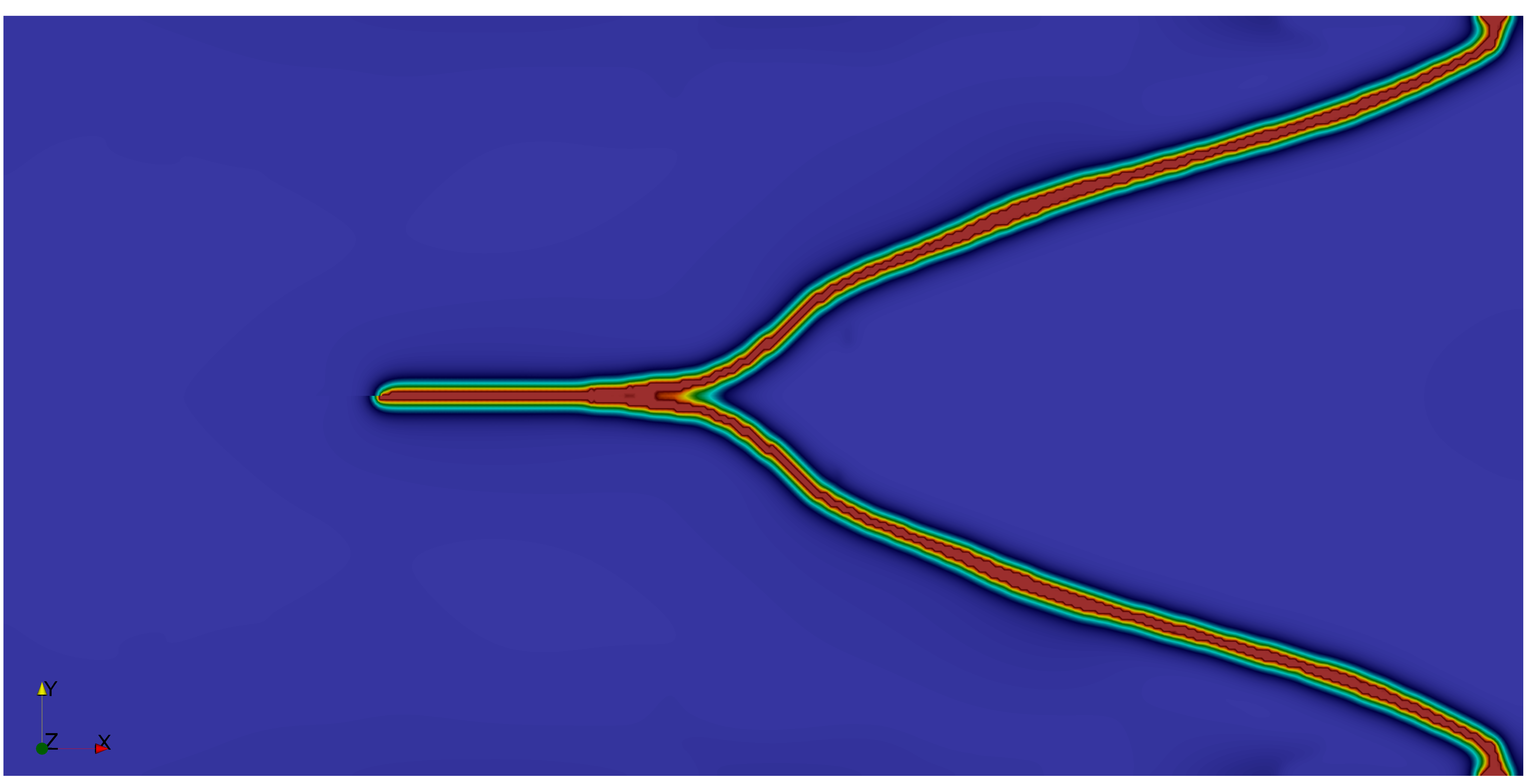}}
    \caption{Phase field,  $t =25 \,ms$}
  \end{subfigure}\\
  \vspace{.3cm}
  \begin{subfigure}[b]{0.85\linewidth}
    \centering%
    {\includegraphics[height=5.5cm,trim={0 0 0 0},clip]{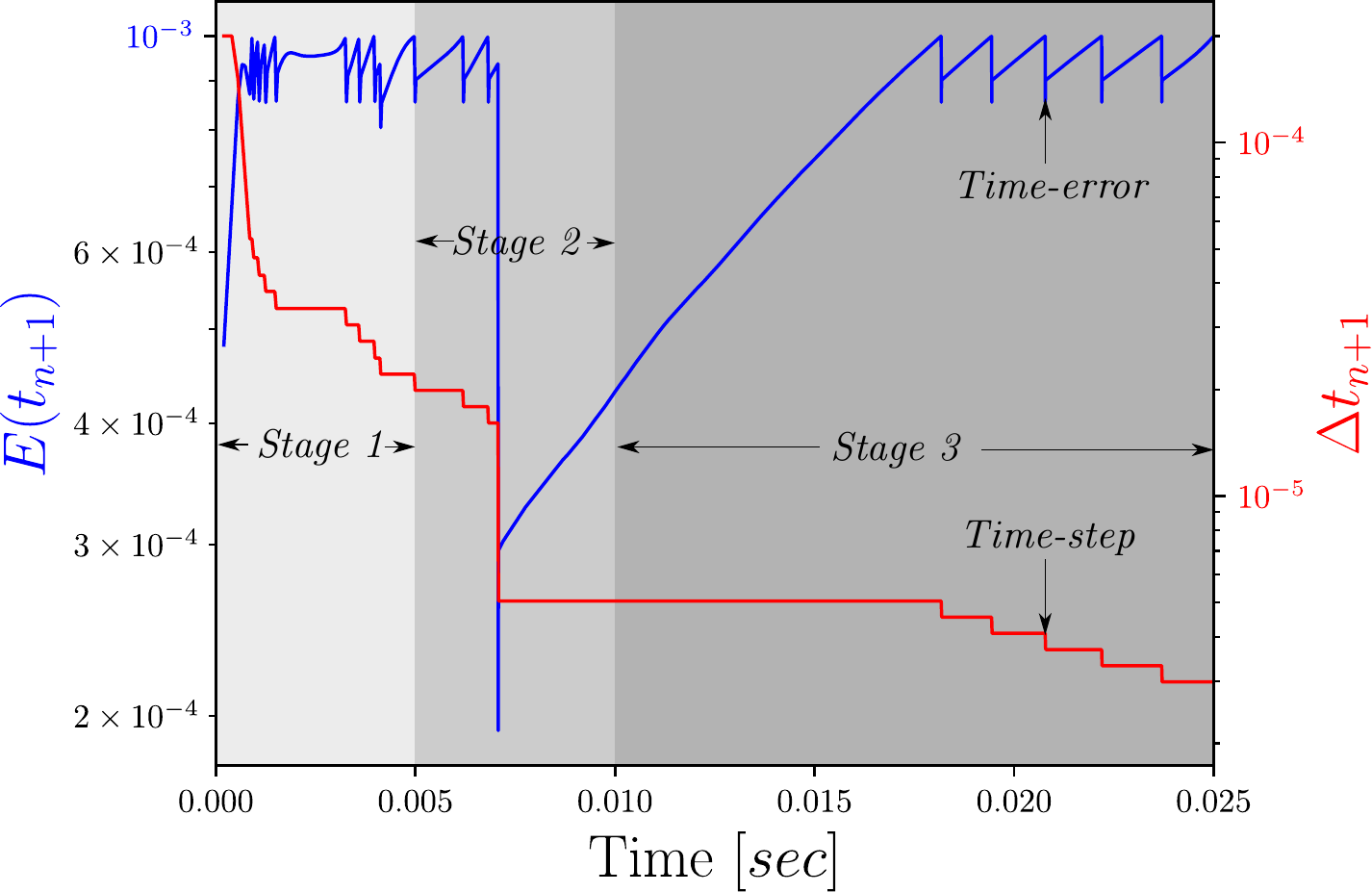}}
    \caption{Error $E \left( t_{n+1} \right)$ \& time-step size $\Delta t$ evolution}
  \end{subfigure}
  \caption{Dynamic fracture branching: time adaptivity based on error control}
  \label{Fig:PhaseFieldTimeAdap}
\end{figure}

Figure~\ref{Fig:PhaseFieldTimeAdapWO}~(a) shows the final crack pattern when we control $\Delta t$ by iteration number in the staggered solution scheme. In particular, we set the staggered iteration number threshold to $10$, reducing the time-step size to satisfy the bound. Figure~\ref{Fig:PhaseFieldTimeAdapWO}~(b) shows the error $E \left( t_{n+1} \right)$ and the time-step-size $\Delta t$ evolution. We divide the last plot into three well-defined stages: stage~1 where the crack nucleation happens, stage~2 where a single crack propagates and finishes when it branches in two, and stage~3 where two independent cracks propagate. The final crack configuration at time $t=25\,ms$ is asymmetric, despite the example is symmetric; the asymmetry is due to the error accumulation during the time-marching process. The most significant increase in the error evolution occurs during stage~1. This simple test shows that the time-step control guided by iteration counts can lead to unreliable results.

 In contrast, Figure \ref{Fig:PhaseFieldTimeAdap} shows the resulting phase-field when we use the error-based approach of Section~\ref{sec:adaptive} to adapt the time-step size. We set $tol_{max} = 10^{-3}$. Figures~\ref{Fig:PhaseFieldTimeAdap}~(a)-(c) display three snapshots that correspond to the phase-field solution on the finest mesh at each stage for times $t=5\,ms$, $t=10\,ms$, and $t=25\,ms$, respectively. Figure \ref{Fig:PhaseFieldTimeAdap}~(d) shows the error and time-step-size evolution. The figure shows that most of the error accumulation occurs during the first stage, which results in a significant time-step size reduction during the first $10\,ms$ of the simulation. Beyond this point, the time-step size remains unchanged until the simulation end, when the crack pathway is already traced. This simulation demonstrates that using an error-based time-adaptive approach delivers reliable results where the final phase-field path is symmetric.


\begin{figure}[h!]
  \begin{subfigure}[b]{0.495\linewidth}
    \centering%
    {\includegraphics[height=5.0cm,trim={0cm 0cm0 0cm 0cm},clip]{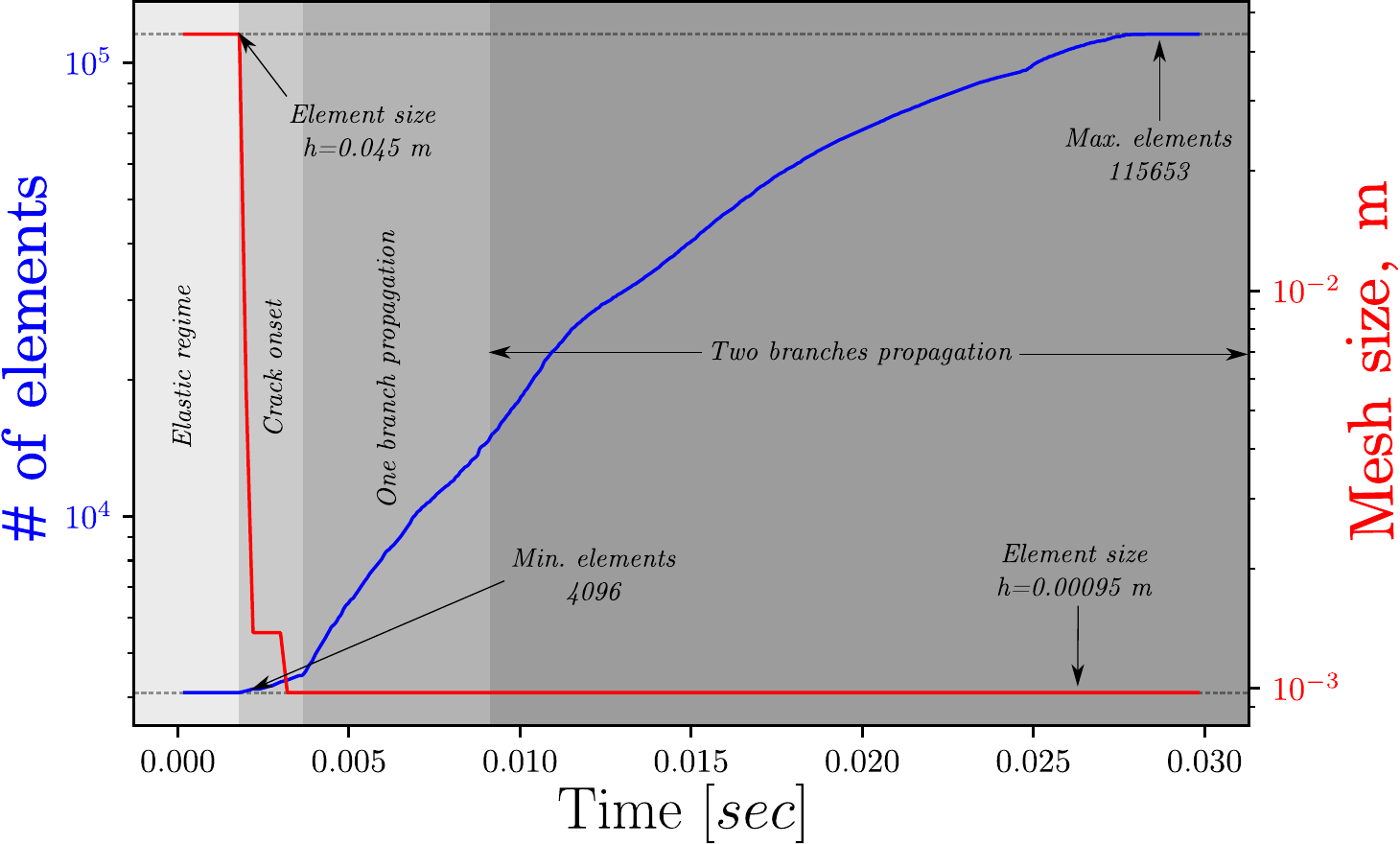}}
    \caption{Element number \& mesh size evolution}
  \end{subfigure}
  \begin{subfigure}[b]{0.495\linewidth}
    \centering%
    {\includegraphics[height=5.0cm,trim={0cm 0cm0 0cm 0cm},clip]{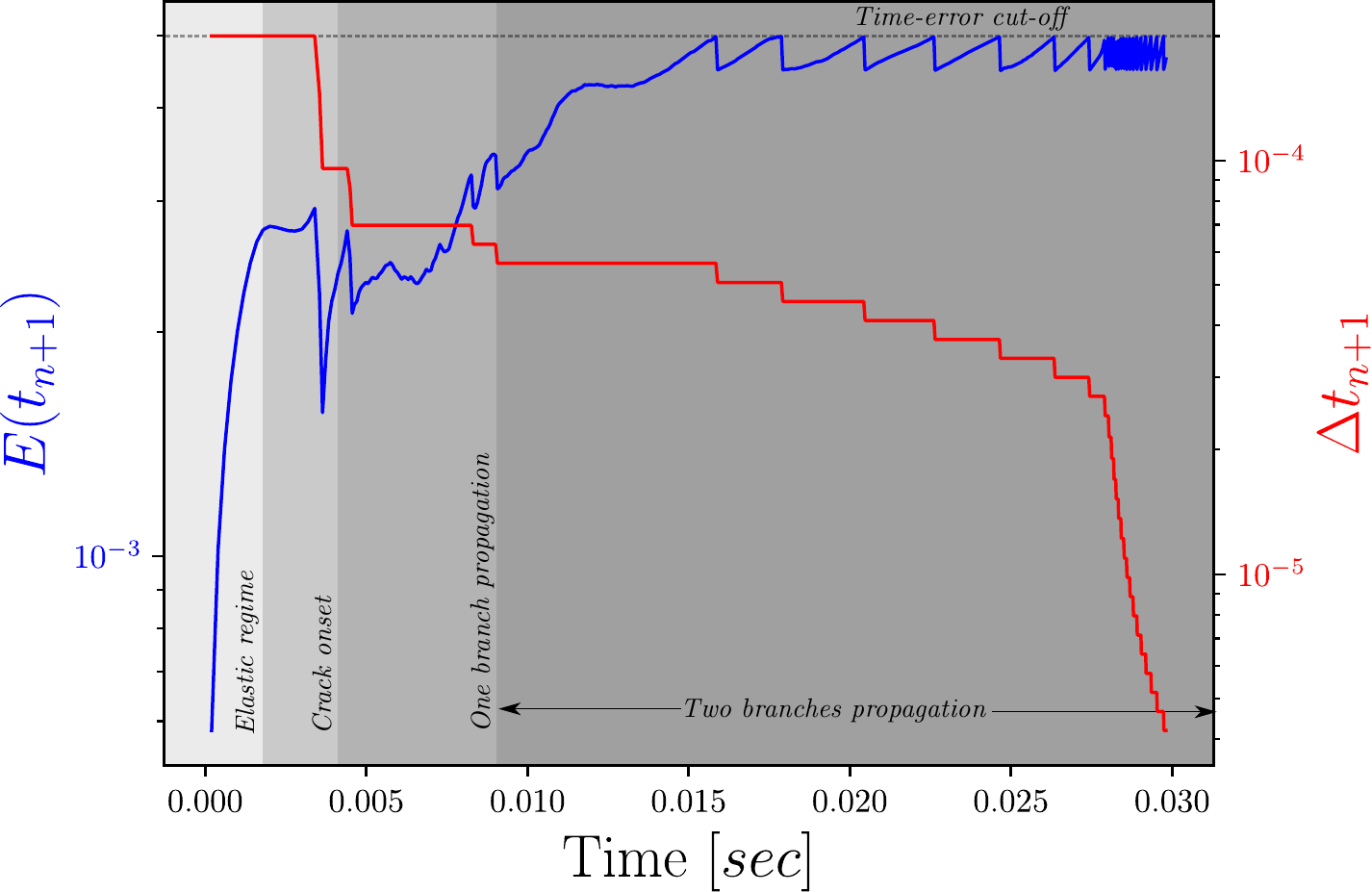}}
    \caption{Error $E \left( t_{n+1} \right)$ \& time-step size $\Delta t$ evolution}
  \end{subfigure}
  \caption{Space-and-time adaptivity}
  \label{Fig:Nelement}
\end{figure}

\subsection{Crack branching with space-and-time adaptivity}

\begin{figure}[h!]
  \centering%
  {\includegraphics[height=5.5cm,trim={0cm 0cm0 0cm 0cm},clip]{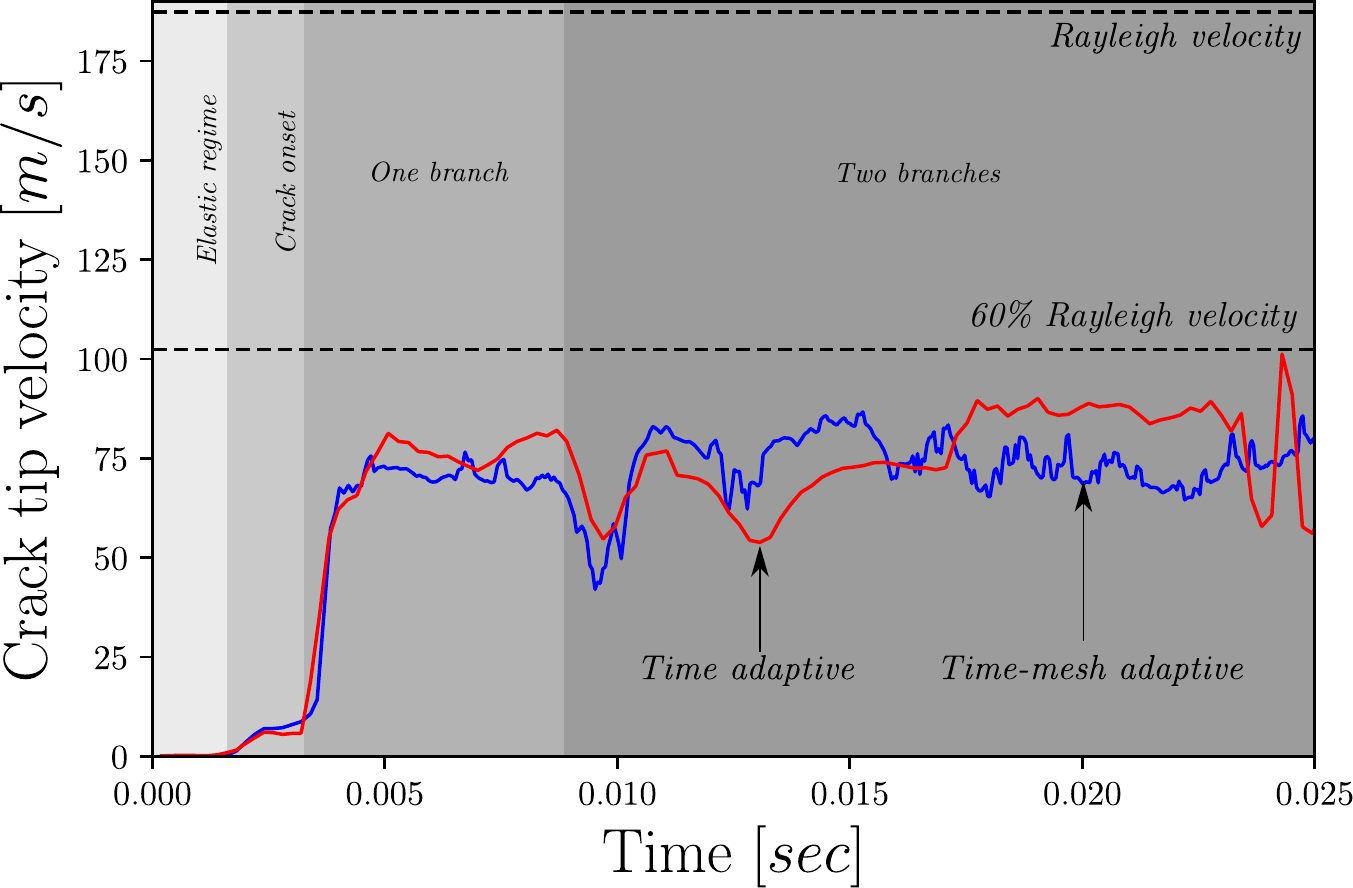}}
  \caption{Crack tip velocity. time adaptivity versus space-and-time adaptivity}
  \label{Fig:Cracktipvel}
\end{figure}

In this numerical experiment, we use our space-and-time adaptive scheme to simulate the problem that Figure~\ref{Fig:BranchingSetup} shows. The initial regular mesh has $4,096$ elements with a mesh size of $45 \, mm$, setting the localization length to $\ell = 5 \, mm$ while we set the maximum tolerance for the time integrator to $tol_{max}=5\times10^{-3}$. Figure~\ref{Fig:Nelement}~(a) shows the time evolution of the number of elements (blue) and the minimum mesh size (red). The final mesh has a minimum element size of $0.95 \, mm$ with a total number of elements of $115,653$, fewer than half the number we used in the previous example. Figure~\ref{Fig:Nelement}~(b) shows the temporal evolution of the error in time $E \left( t_{n+1} \right)$ (blue) and the time-step size $\Delta t_{n+1}$ (red). We divide the evolution into four well-defined stages. The first stage represents the elastic regime, where a coarse mesh can appropriately describe the evolution. The second stage contains the crack onset, where the algorithm refines the mesh to capture the tip notch evolution, wherein we need a smaller element size. We introduce a refinement cut-off in this case to avoid unnecessarily small elements. The third stage represents a single branch propagation, where the element number increases faster, while the time-step size also decreases considerably. The fourth stage depicts the crack branching and the propagation of two fractures. In this last stage, the algorithm deploys the maximum number of elements.

\begin{figure}[h!]
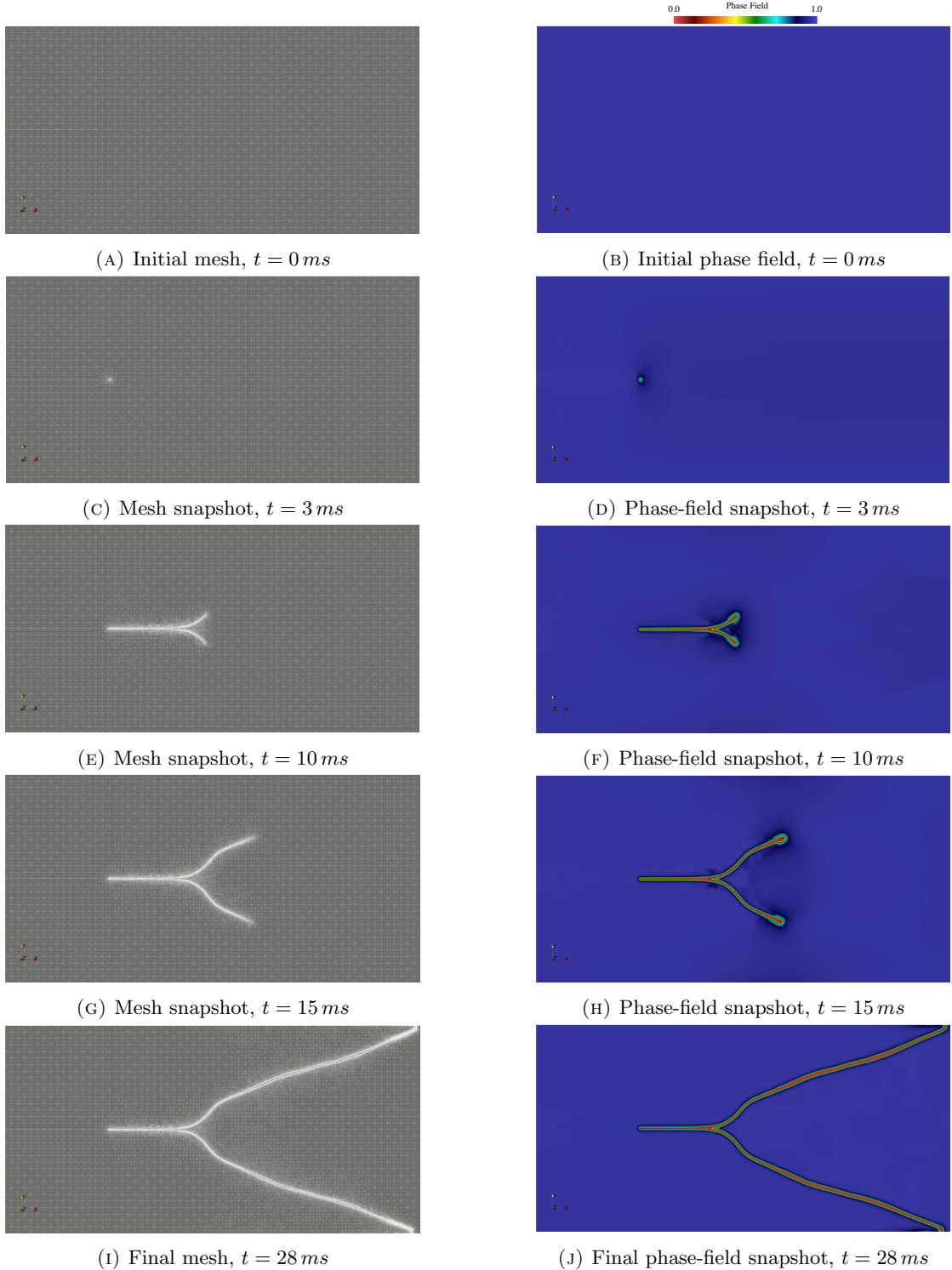

  \centering%
  \begin{subfigure}[b]{0.495\linewidth}
    \centering%
    {\includegraphics[width=6.5cm,trim={0cm 0cm0 0cm 0cm},clip]{Mesh0000s.pdf}}
    \caption{Initial mesh, $t = 0 \, ms$}
  \end{subfigure}
  \begin{subfigure}[b]{0.495\linewidth}
    \centering%
    {\includegraphics[width=6.5cm,trim={0cm 0cm0 0cm 0cm},clip]{PhaseField0000s.pdf}}
    \caption{Initial phase field, $t = 0 \, ms$}
  \end{subfigure}
  \begin{subfigure}[b]{0.495\linewidth}
    \centering%
    {\includegraphics[width=6.5cm,trim={0cm 0cm0 0cm 0cm},clip]{Mesh0003s.pdf}}
    \caption{Mesh snapshot, $t = 3 \, ms$}
  \end{subfigure}
  \begin{subfigure}[b]{0.495\linewidth}
    \centering%
    {\includegraphics[width=6.5cm,trim={0cm 0cm0 0cm 0cm},clip]{PhaseField0003s.pdf}}
    \caption{Phase-field snapshot, $t = 3 \, ms$}
  \end{subfigure}
  \begin{subfigure}[b]{0.495\linewidth}
    \centering%
    {\includegraphics[width=6.5cm,trim={0cm 0cm0 0cm 0cm},clip]{Mesh0011s.pdf}}
    \caption{Mesh snapshot, $t = 10 \, ms$}
  \end{subfigure}
  \begin{subfigure}[b]{0.495\linewidth}
    \centering%
    {\includegraphics[width=6.5cm,trim={0cm 0cm0 0cm 0cm},clip]{PhaseField0011s.pdf}}
    \caption{Phase-field snapshot, $t = 10 \, ms$}
  \end{subfigure}
  \begin{subfigure}[b]{0.495\linewidth}
    \centering%
    {\includegraphics[width=6.5cm,trim={0cm 0cm0 0cm 0cm},clip]{Mesh0015s.pdf}}
    \caption{Mesh snapshot, $t = 15 \, ms$}
  \end{subfigure}
  \begin{subfigure}[b]{0.495\linewidth}
    \centering%
    {\includegraphics[width=6.5cm,trim={0cm 0cm0 0cm 0cm},clip]{PhaseField0015s.pdf}}
    \caption{Phase-field snapshot, $t = 15 \, ms$}
  \end{subfigure}
  \begin{subfigure}[b]{0.495\linewidth}
    \centering%
    {\includegraphics[width=6.5cm,trim={0cm 0cm0 0cm 0cm},clip]{Mesh0029s.pdf}}
    \caption{Final mesh, $t = 28\, ms$}
  \end{subfigure}
  \begin{subfigure}[b]{0.495\linewidth}
    \centering%
    {\includegraphics[width=6.5cm,trim={0cm 0cm0 0cm 0cm},clip]{PhaseField0029s.pdf}}
    \caption{Final phase-field  snapshot, $t = 28\, ms$}
  \end{subfigure}
  \caption{Space-and-time adaptive simulation of branching fracture
    (cubic degradation function): Mesh \& phase-field evolution}
  \label{Fig:TimeMesh}
\end{figure}

Figure~\ref{Fig:Cracktipvel} compares the crack tip velocity between the time adaptive example presented in the previous section and the space-and-time adaptive case. The figures show similar results with four well-defined sections: elastic regime, crack nucleation, and propagation of one branch and two branches. Also, the crack propagation is around the $60\%$ of the Rayleigh velocity of the solid body.

Furthermore, Figure~\ref{Fig:TimeMesh} shows a snapshot timeline of the simulation results, with the phase field and the computed mesh in each case. For time $t = 3 \, ms$, Figures~\ref{Fig:TimeMesh}~(c) and~(d) show the crack nucleation, while at $t = 10 \, ms$, Figures~\ref{Fig:TimeMesh}~(e) and~(f) show the crack branching.  Finally, Figures~\ref{Fig:TimeMesh}~(g) to~(j) show the final stages of the problem, where two cracks propagate producing a symmetric crack profile by properly resolving the dynamics locally at the crack tips.


\begin{figure}
\begin{subfigure}[b]{0.495\linewidth}
\centering%
{\includegraphics[height=4.15cm,trim={0cm 0cm0 0cm 0cm},clip]{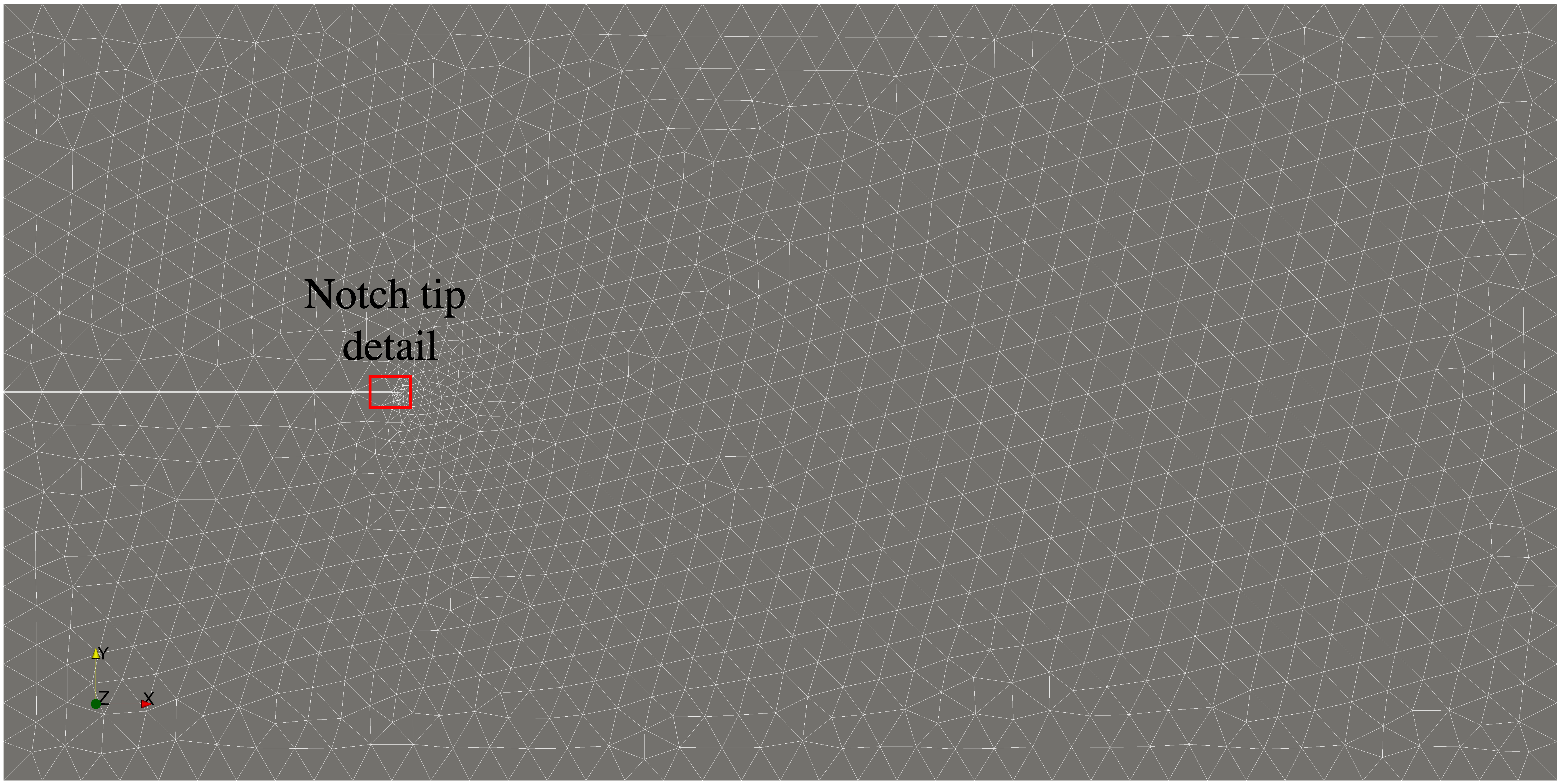}}
\caption{Initial mesh (overview)}
\end{subfigure}
\begin{subfigure}[b]{0.495\linewidth}
\centering%
{\includegraphics[height=4.15cm,trim={27cm -1.5cm 27cm -0.75cm},clip]{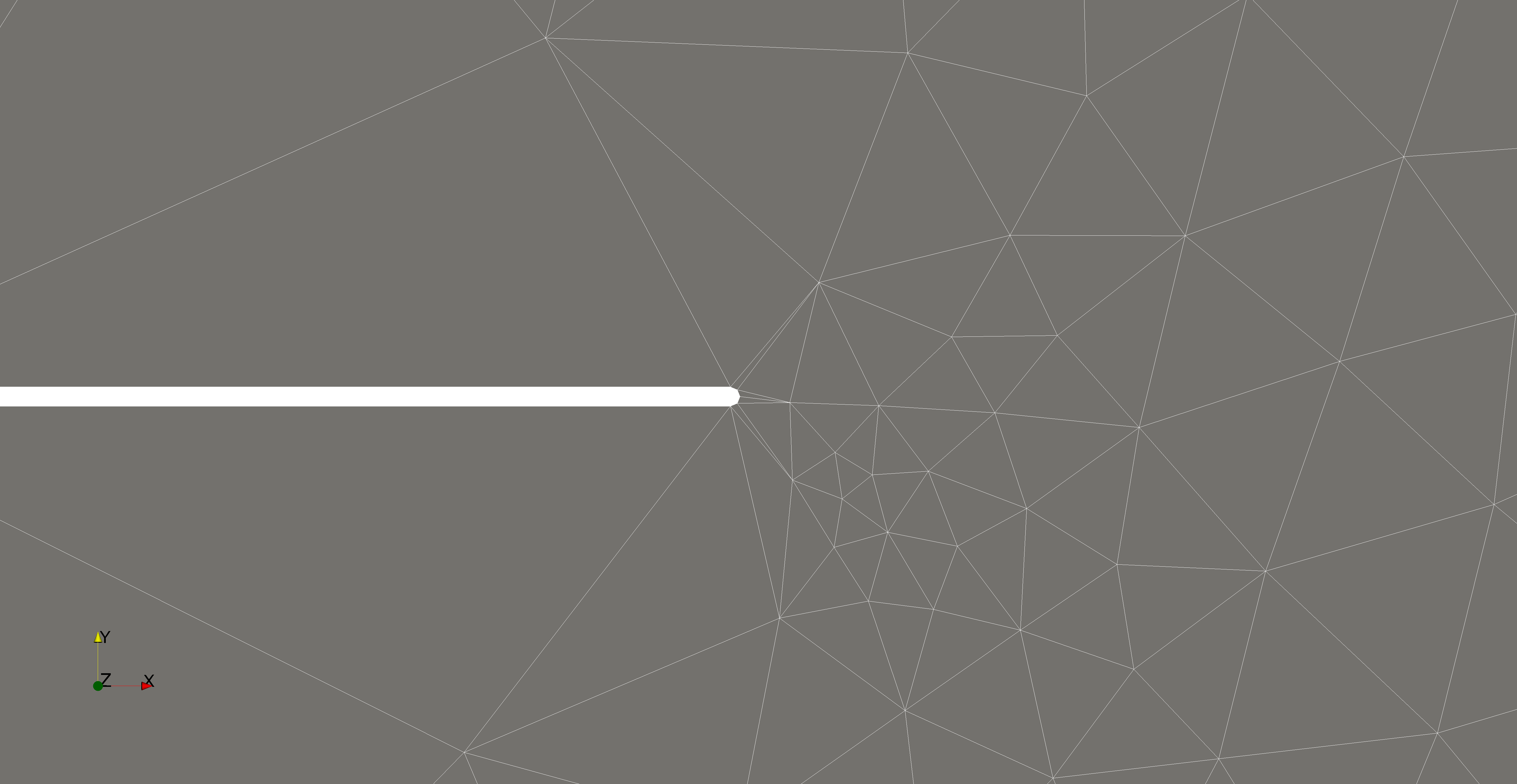}}
\caption{Notch tip detail}
\end{subfigure}
\caption{Unstructured initial mesh}
\label{Fig:irregularinitialmesh}
\end{figure}

\begin{figure}
\begin{subfigure}[b]{0.495\linewidth}
\centering%
{\includegraphics[height=4.5cm,trim={0cm 0cm0 0cm 0cm},clip]{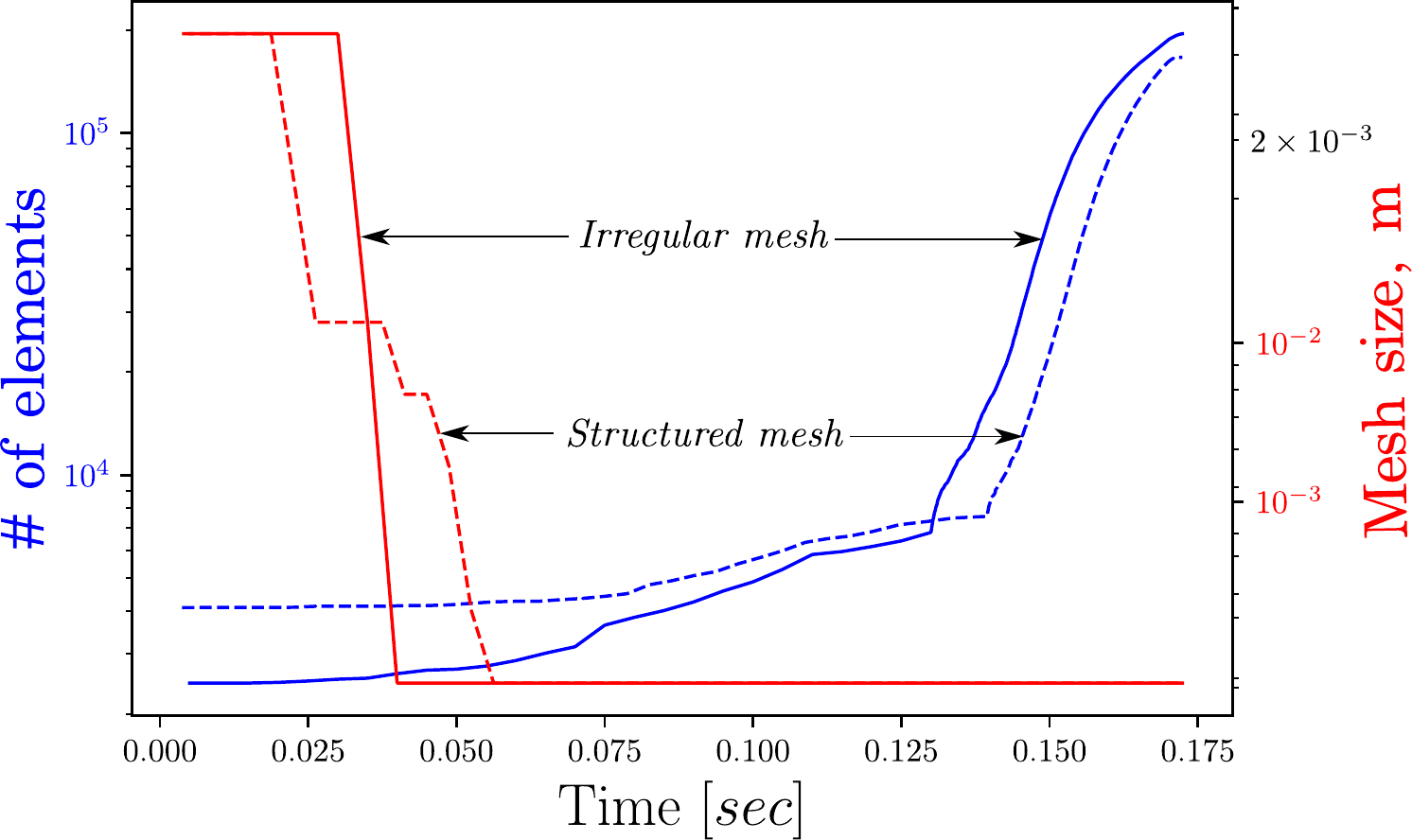}}
\caption{Element number \& mesh size evolution}
\end{subfigure}
\begin{subfigure}[b]{0.495\linewidth}
\centering%
{\includegraphics[height=4.5cm,trim={0cm 0cm 0cm 0cm},clip]{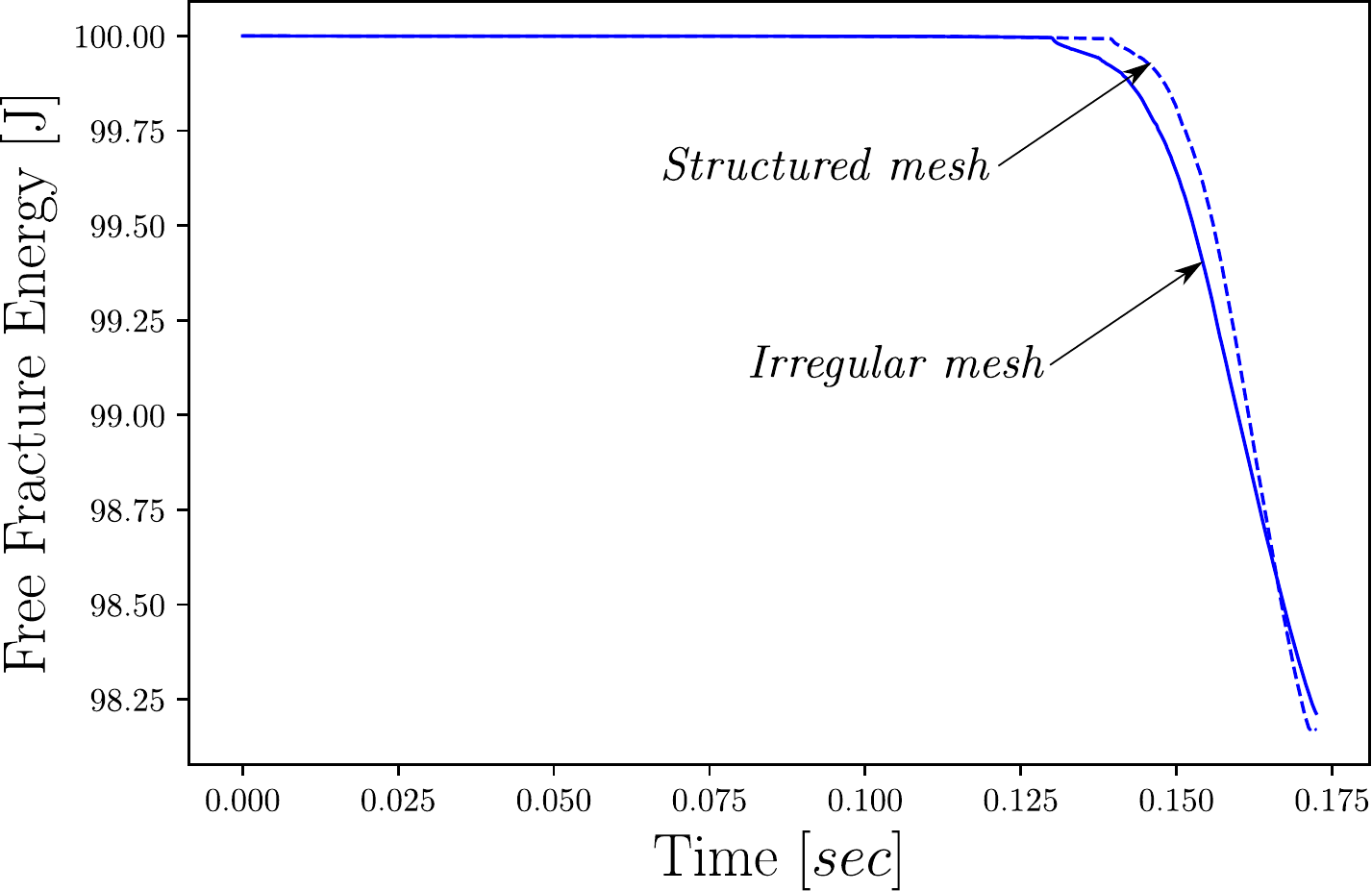}}
\caption{Free fracture energy evolution}
\end{subfigure}
\caption{Structured versus unstructured (highly irregular) meshes}
\label{Fig:Dissipatedenergy}
\end{figure}

\begin{figure}
\centering%
\begin{subfigure}[b]{0.495\linewidth}
\centering%
{\includegraphics[width=7.5cm,trim={0cm 0cm 0cm 0cm},clip]{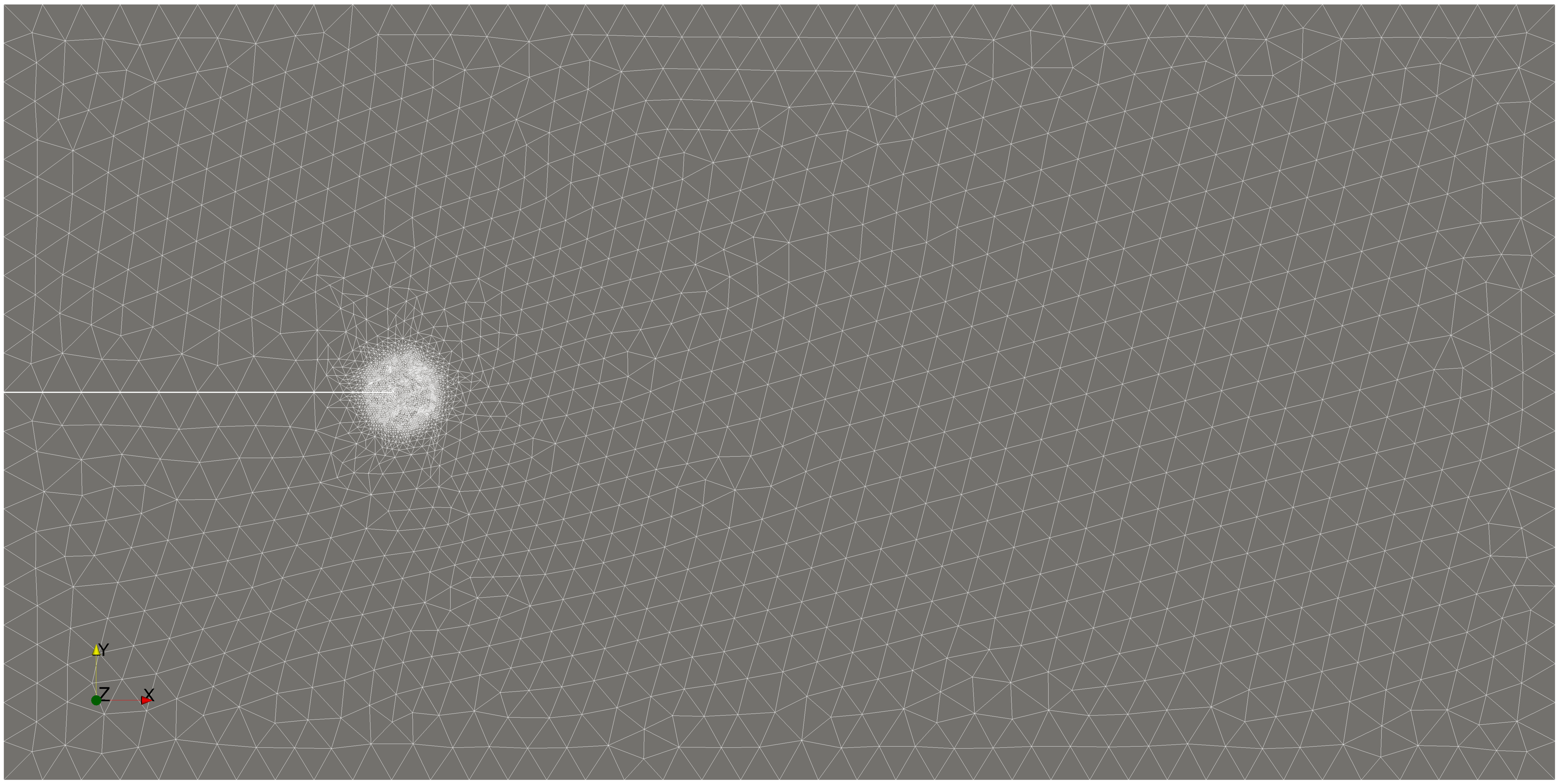}}
\caption{Unstructured mesh, $t = 130 \, ms$}
\end{subfigure}
\begin{subfigure}[b]{0.495\linewidth}
\centering%
{\includegraphics[width=7.5cm,trim={0cm 0cm 0cm 0cm},clip]{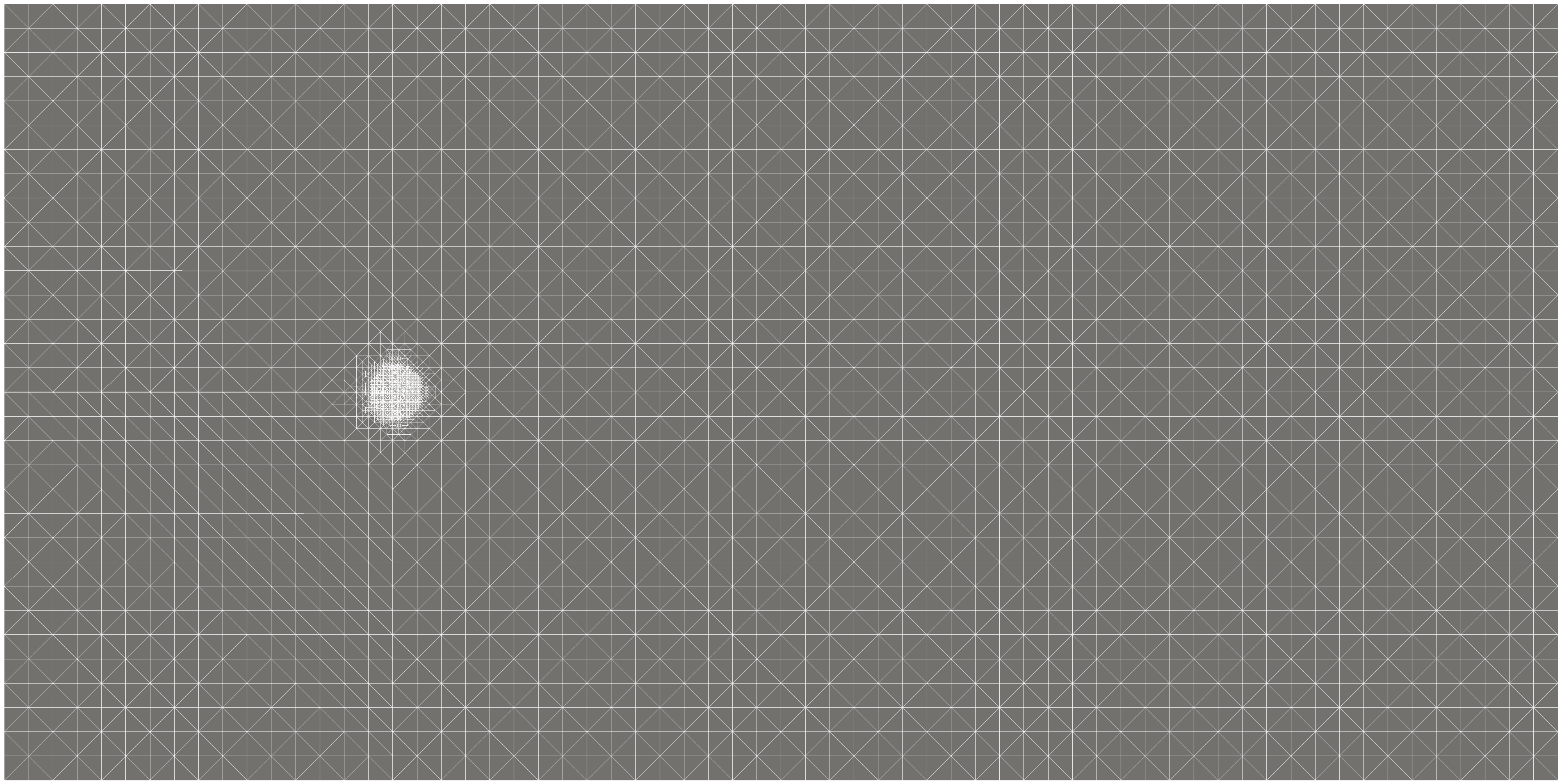}}
\caption{Structured mesh for $t = 130 \,ms$}
\end{subfigure}
\begin{subfigure}[b]{0.495\linewidth}
\centering%
{\includegraphics[width=7.5cm,trim={0cm 0cm 0cm 0cm},clip]{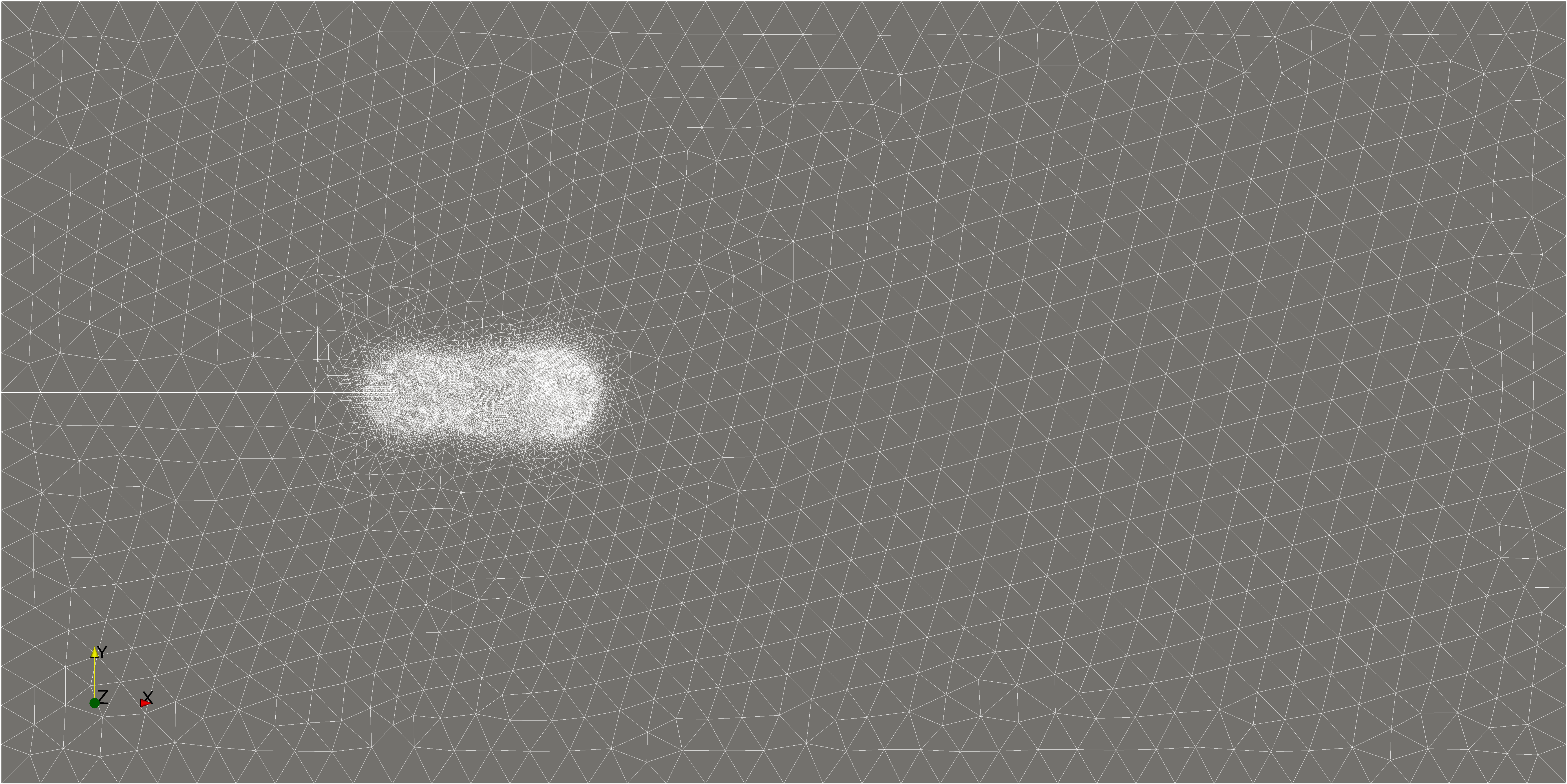}}
\caption{Unstructured mesh, $t = 140 \,ms$}
\end{subfigure}
\begin{subfigure}[b]{0.495\linewidth}
\centering%
{\includegraphics[width=7.5cm,trim={0cm 0cm 0cm 0cm},clip]{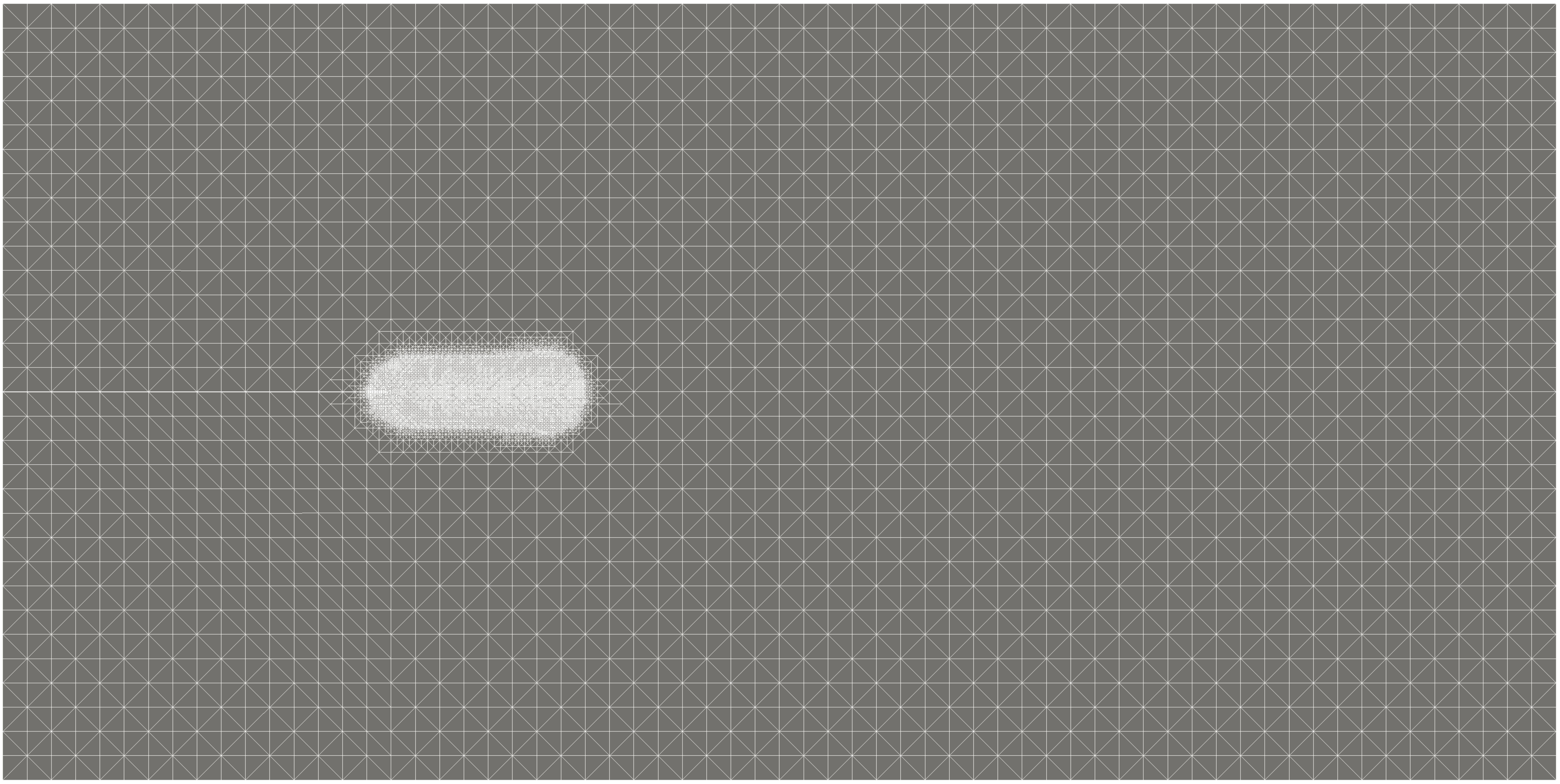}}
\caption{Structured mesh, $t = 140 \, ms$}
\end{subfigure}
\begin{subfigure}[b]{0.495\linewidth}
\centering%
{\includegraphics[width=7.5cm,trim={0cm 0cm 0cm 0cm},clip]{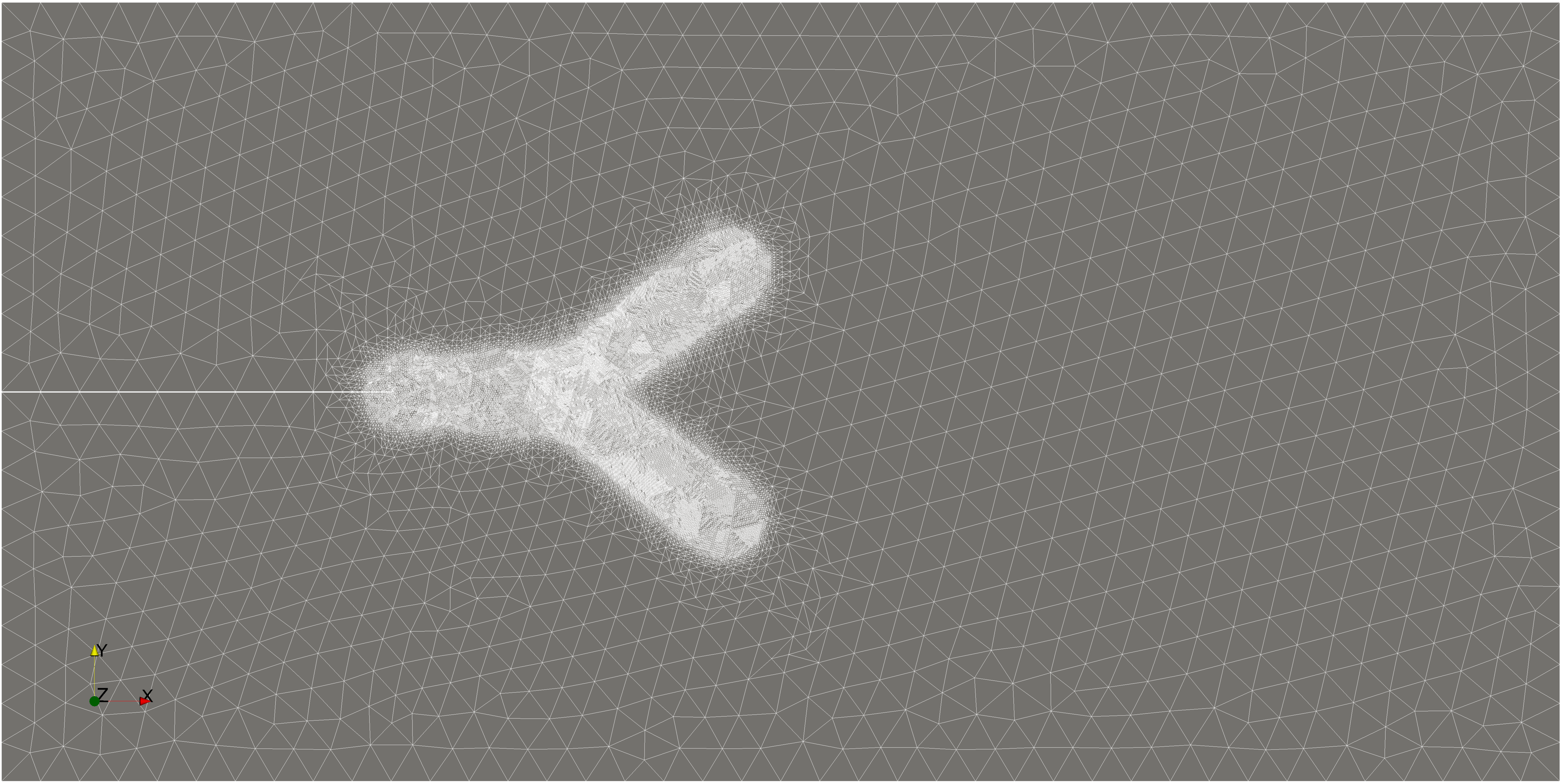}}
\caption{Unstructured mesh, $t = 160 \, ms$}
\end{subfigure}
\begin{subfigure}[b]{0.495\linewidth}
\centering%
{\includegraphics[width=7.5cm,trim={0cm 0cm 0cm 0cm},clip]{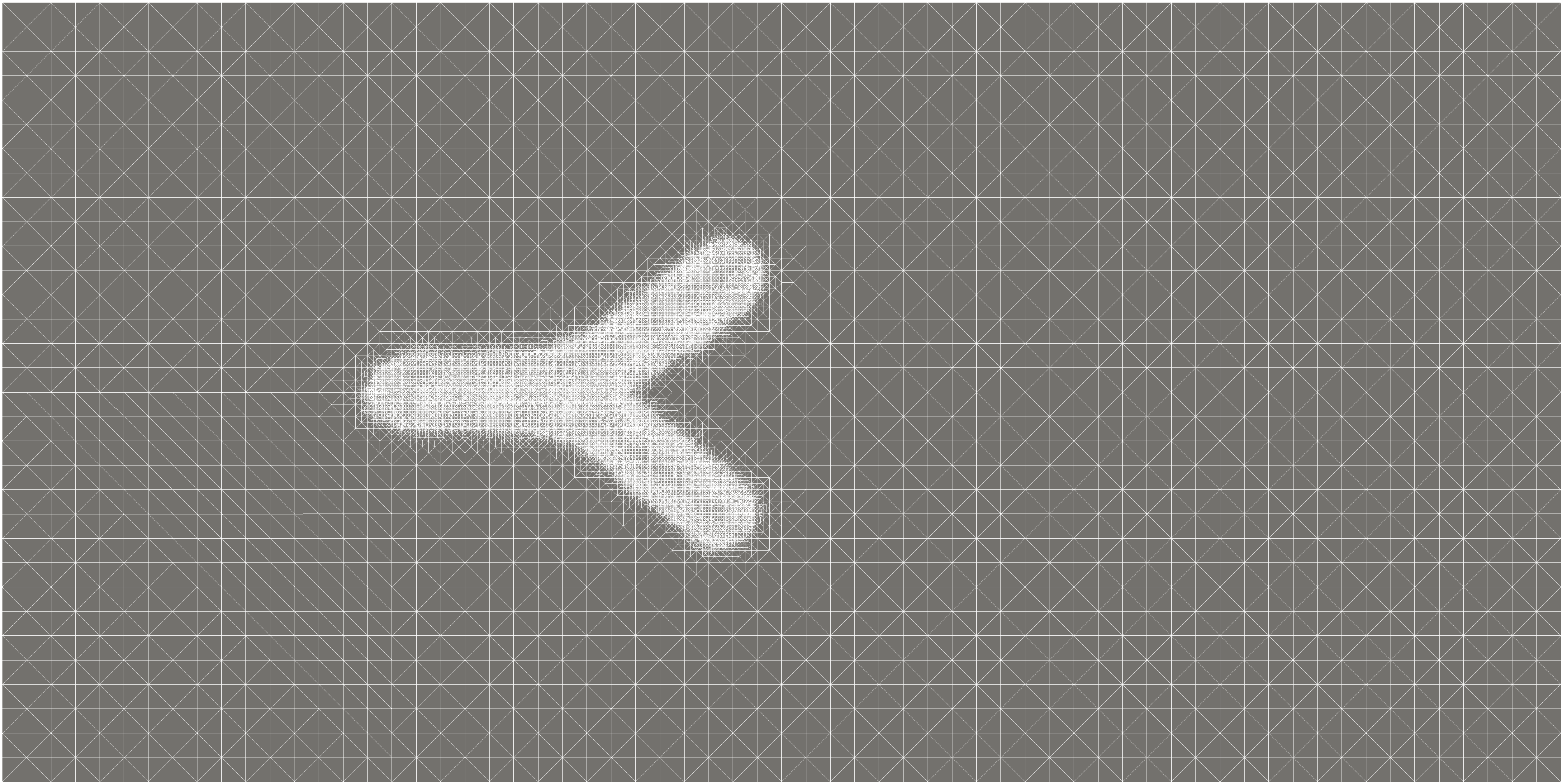}}
\caption{Structured mesh, $t = 160 \, ms$}
\end{subfigure}
\begin{subfigure}[b]{0.495\linewidth}
\centering%
{\includegraphics[width=7.5cm,trim={0cm 0cm 0cm 0cm},clip]{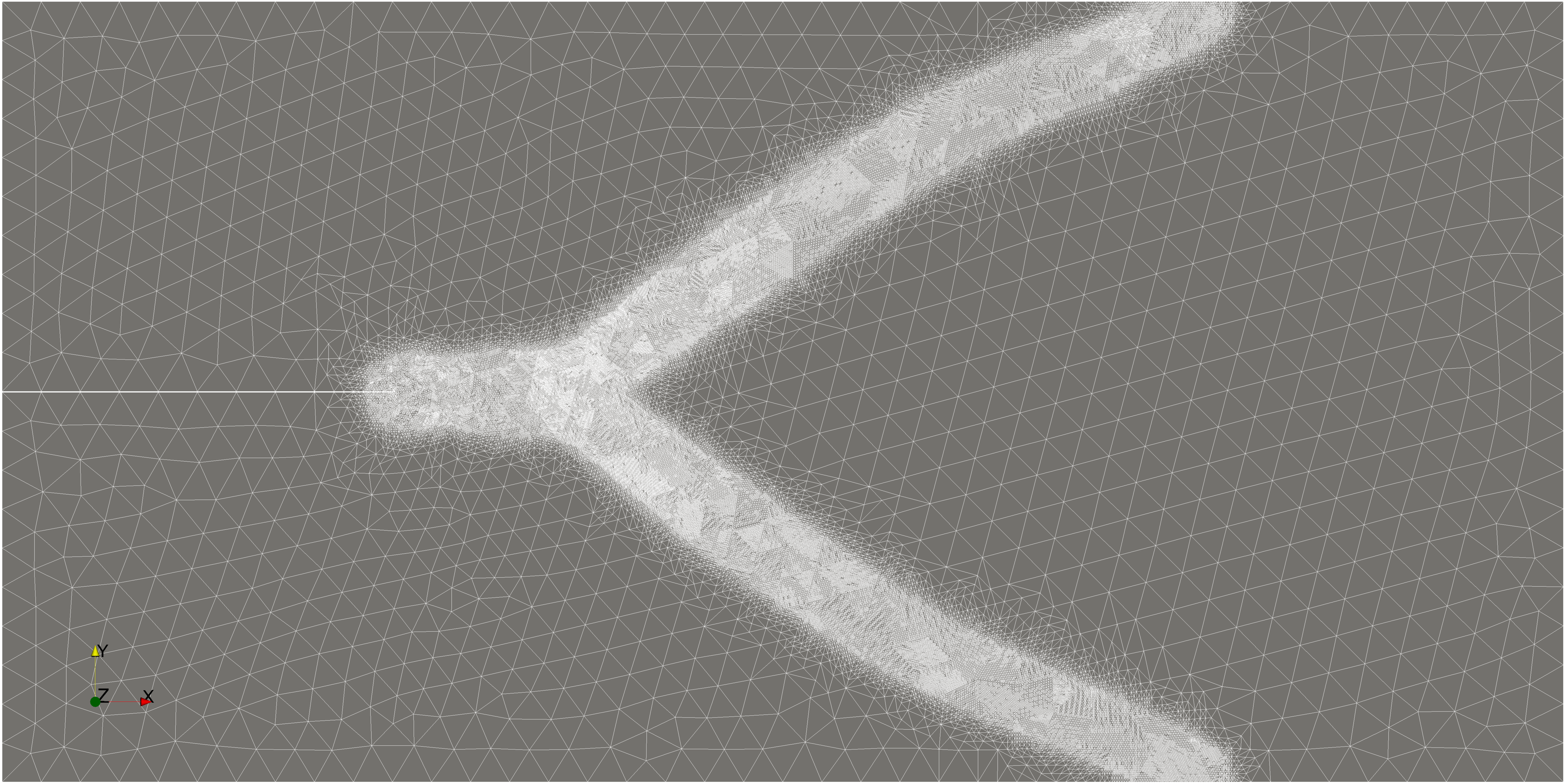}}
\caption{Unstructured mesh, $t = 173 \, ms$}
\end{subfigure}
\begin{subfigure}[b]{0.495\linewidth}
\centering%
{\includegraphics[width=7.5cm,trim={0cm 0cm 0cm 0cm},clip]{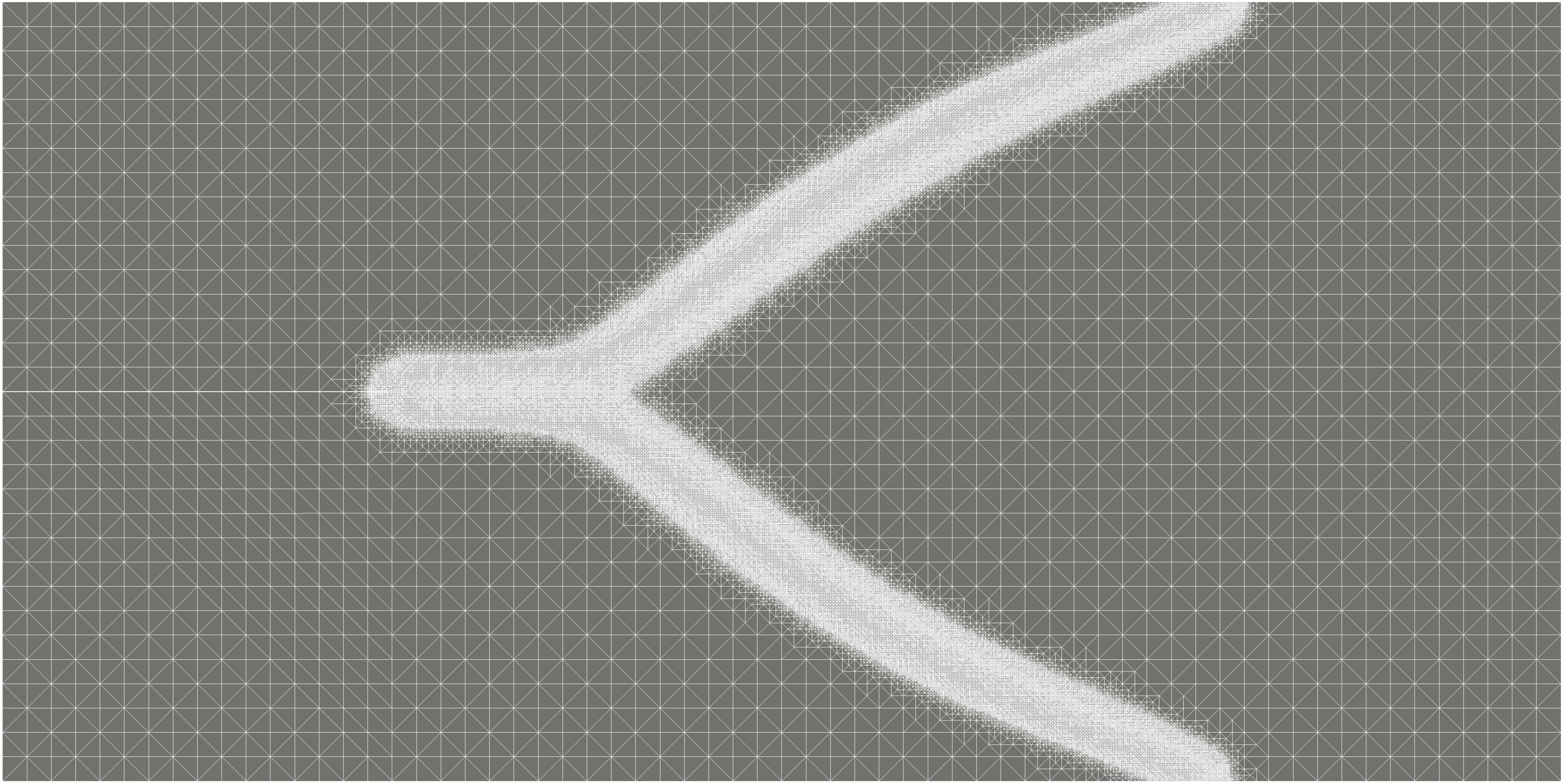}}
\caption{Structured mesh, $t = 173 \, ms$}
\end{subfigure}
  \caption{Space-and-time adaptive simulation of branching fracture
    (cubic degradation function): Mesh evolution}
\label{Fig:TimeMeshirreg}
\end{figure}

\begin{figure}
\centering%
\begin{subfigure}[b]{0.495\linewidth}
\centering%
{\includegraphics[width=7.5cm,trim={0cm 0cm 0cm 0cm},clip]{irrPPt013245.pdf}}
\caption{Unstructured mesh, $t = 130 \, ms$}
\end{subfigure}
\begin{subfigure}[b]{0.495\linewidth}
\centering%
{\includegraphics[width=7.5cm,trim={0cm 0cm 0cm 0cm},clip]{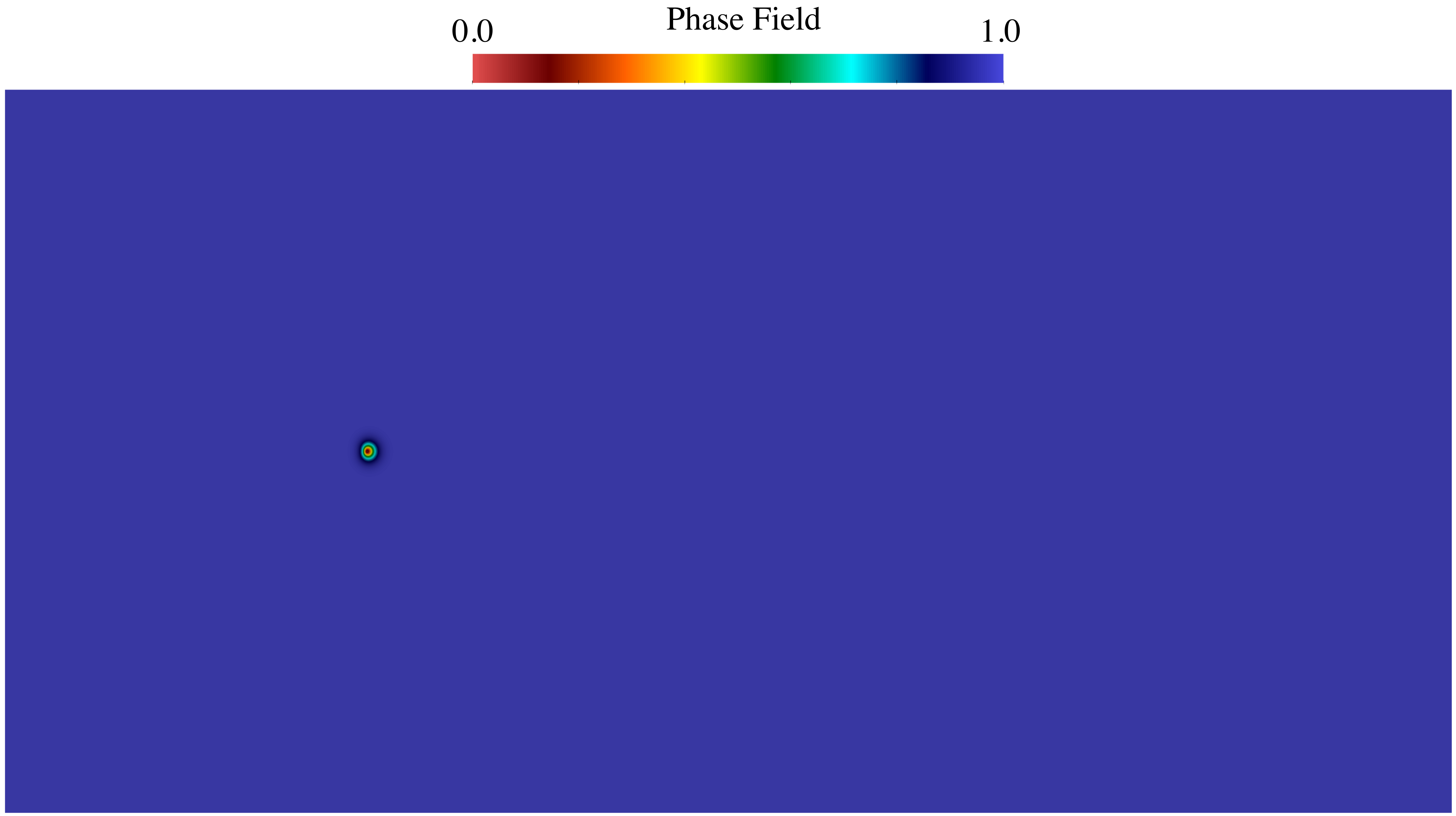}}
\caption{Structured mesh, $t = 130 \, ms$}
\end{subfigure}
\begin{subfigure}[b]{0.495\linewidth}
\centering%
{\includegraphics[width=7.5cm,trim={0cm 0cm 0cm 0cm},clip]{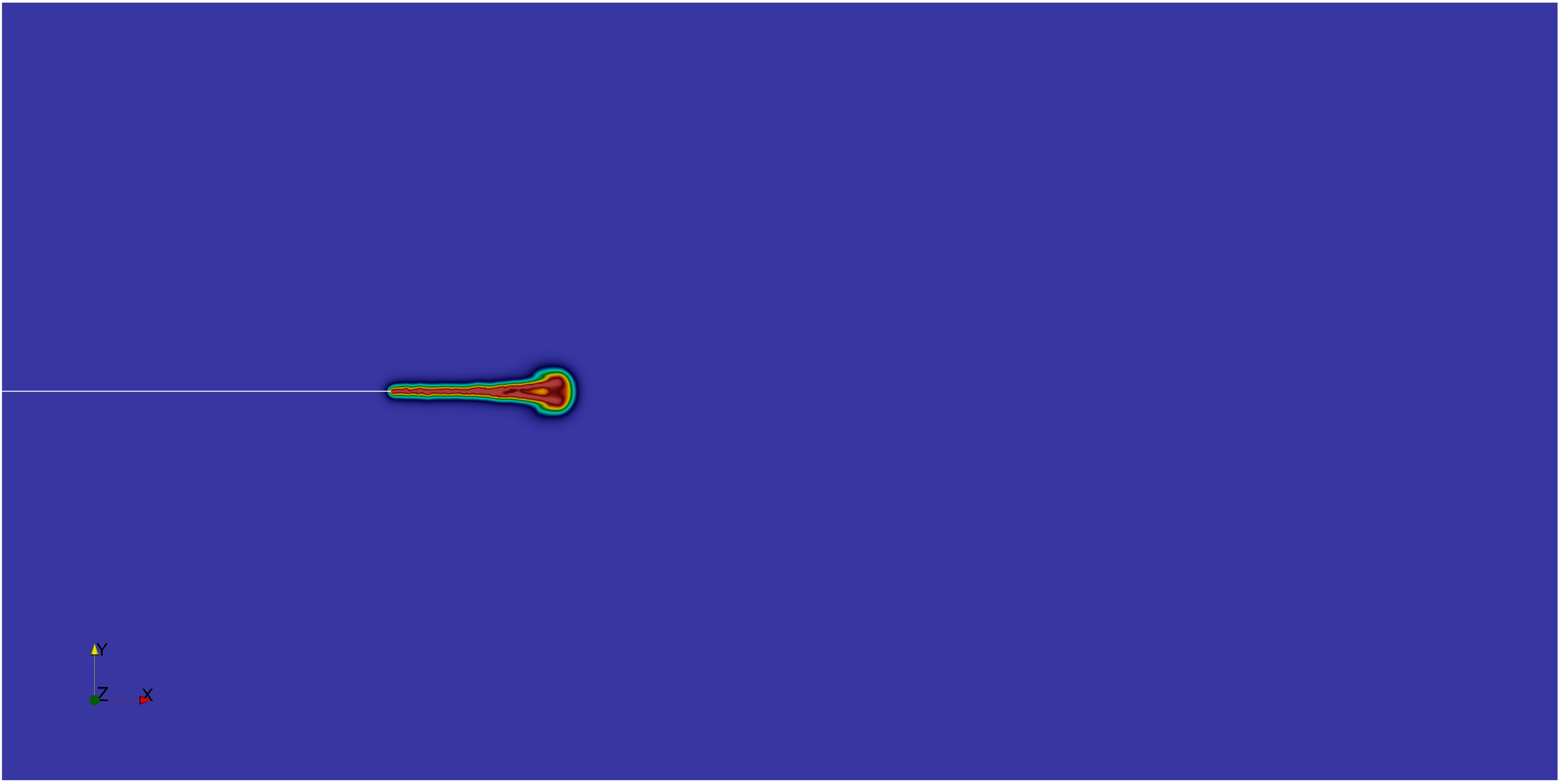}}
\caption{Unstructured mesh, $t = 140 \, ms$}
\end{subfigure}
\begin{subfigure}[b]{0.495\linewidth}
\centering%
{\includegraphics[width=7.5cm,trim={0cm 0cm 0cm 0cm},clip]{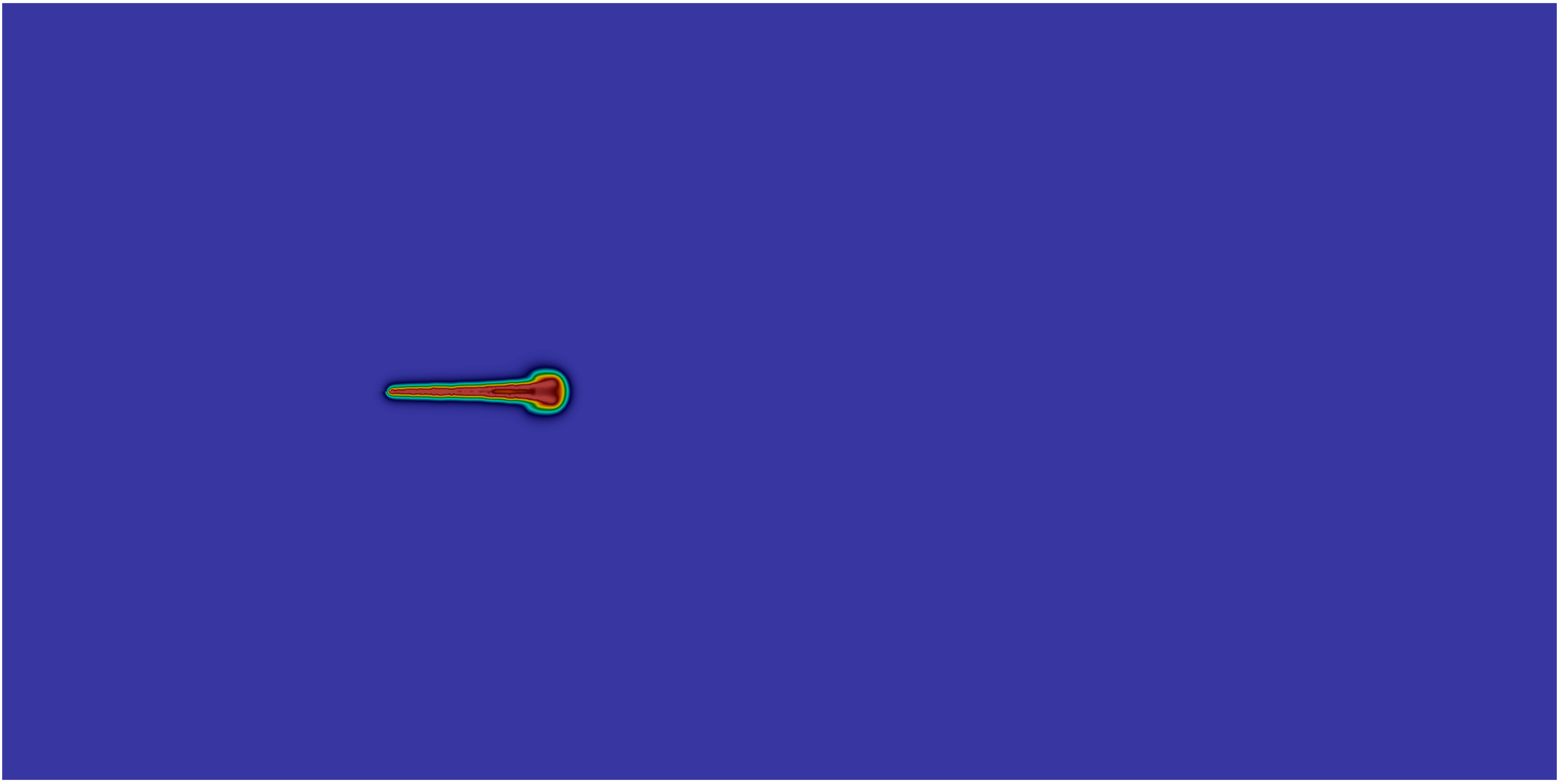}}
\caption{Structured mesh, $t = 140 \, ms$}
\end{subfigure}
\begin{subfigure}[b]{0.495\linewidth}
\centering%
{\includegraphics[width=7.5cm,trim={0cm 0cm 0cm 0cm},clip]{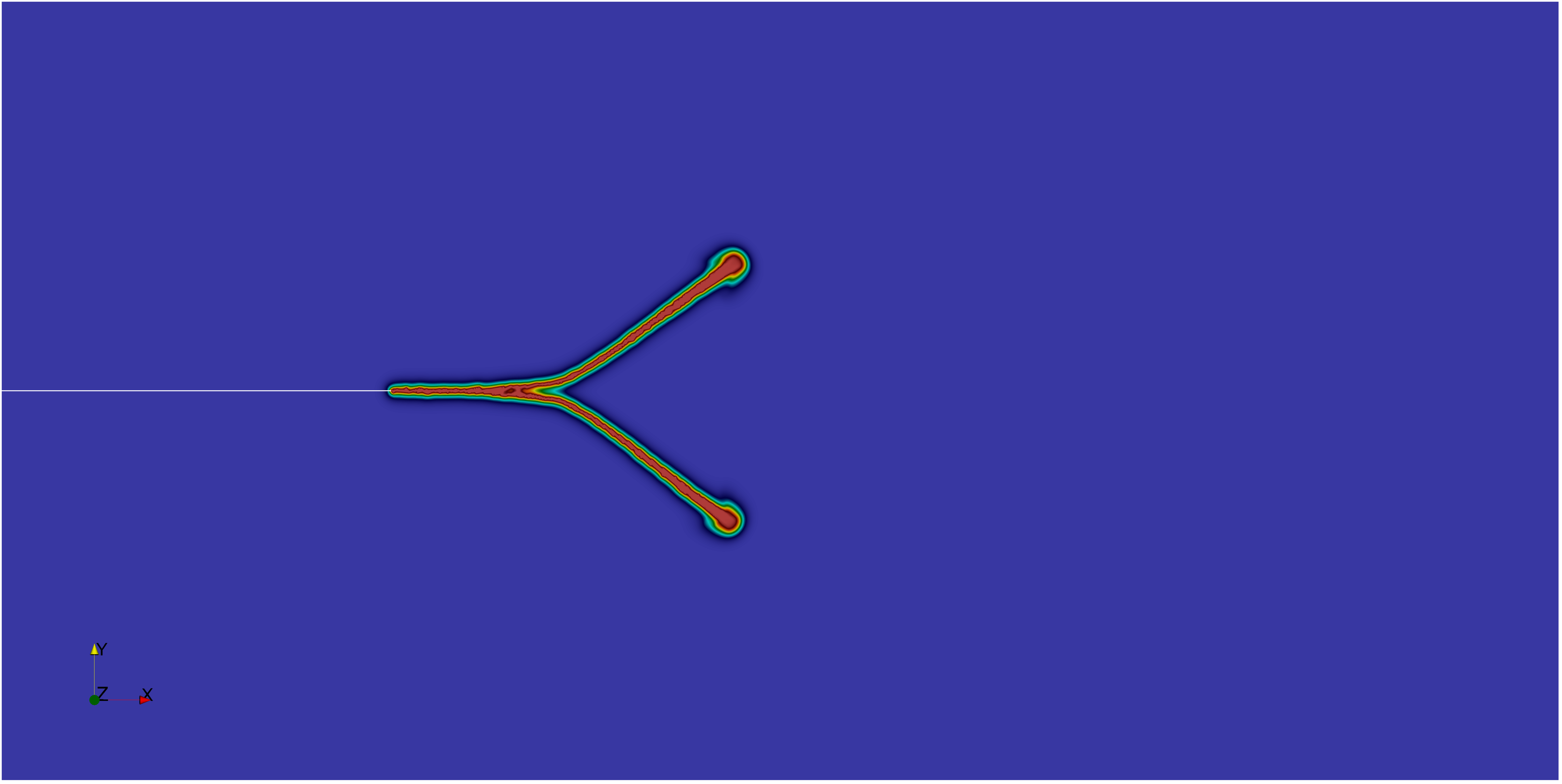}}
\caption{Unstructured mesh, $t = 160 \, ms$}
\end{subfigure}
\begin{subfigure}[b]{0.495\linewidth}
\centering%
{\includegraphics[width=7.5cm,trim={0cm 0cm 0cm 0cm},clip]{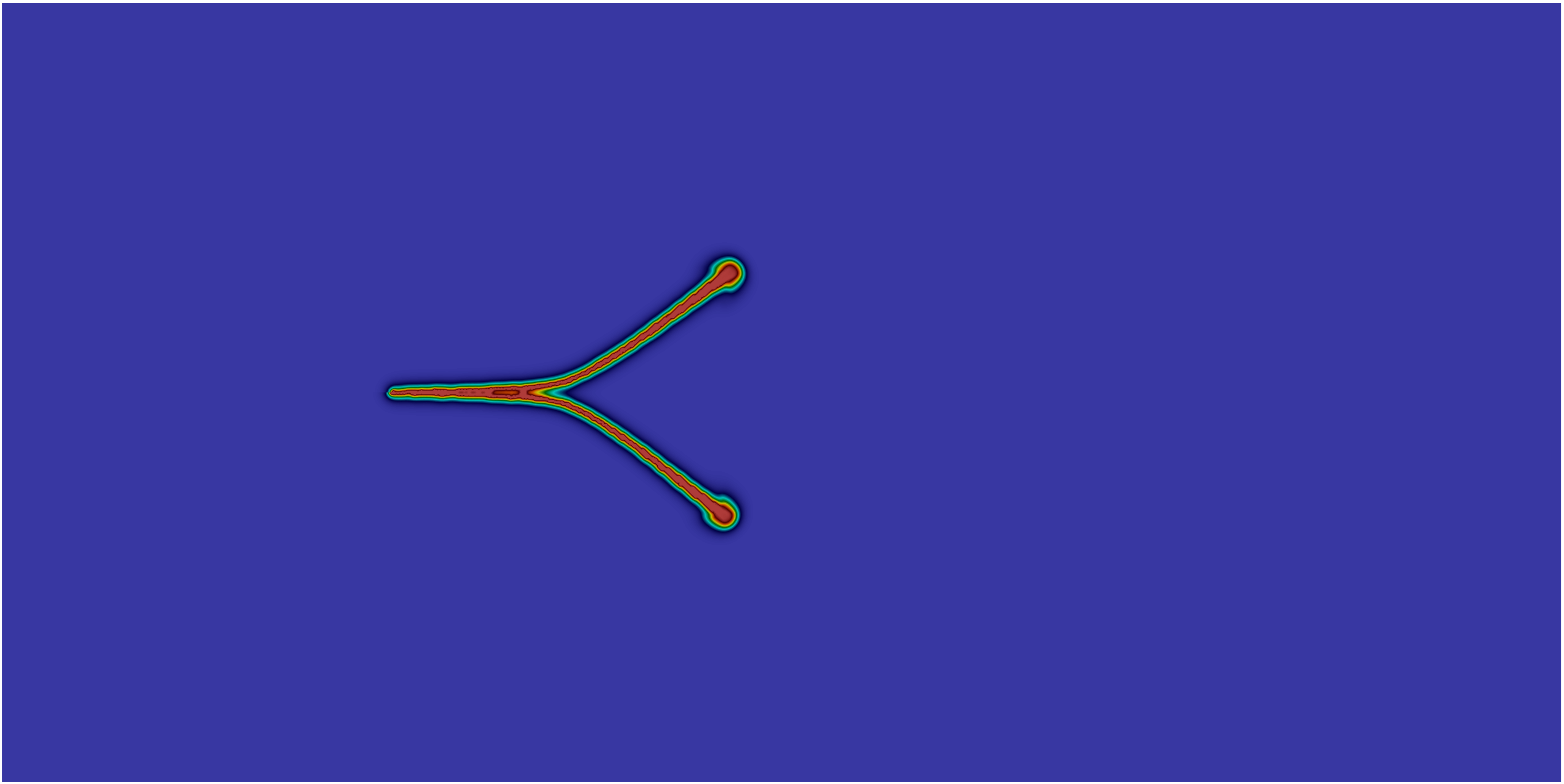}}
\caption{Structured mesh, $t = 160 \, ms$}
\end{subfigure}
\begin{subfigure}[b]{0.495\linewidth}
\centering%
{\includegraphics[width=7.5cm,trim={0cm 0cm 0cm 0cm},clip]{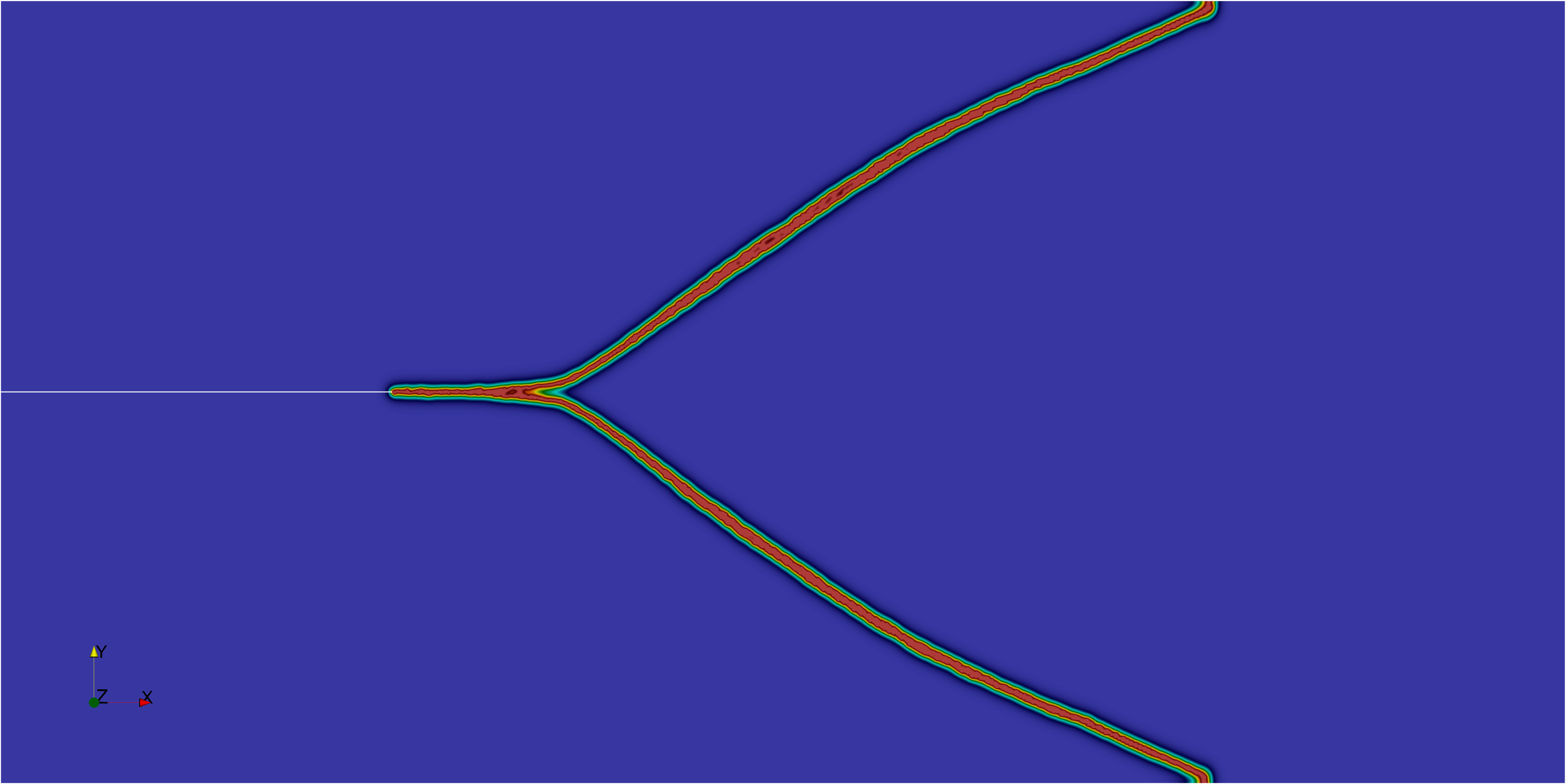}}
\caption{Unstructured mesh, $t = 173 \, ms$}
\end{subfigure}
\begin{subfigure}[b]{0.495\linewidth}
\centering%
{\includegraphics[width=7.5cm,trim={0cm 0cm 0cm 0cm},clip]{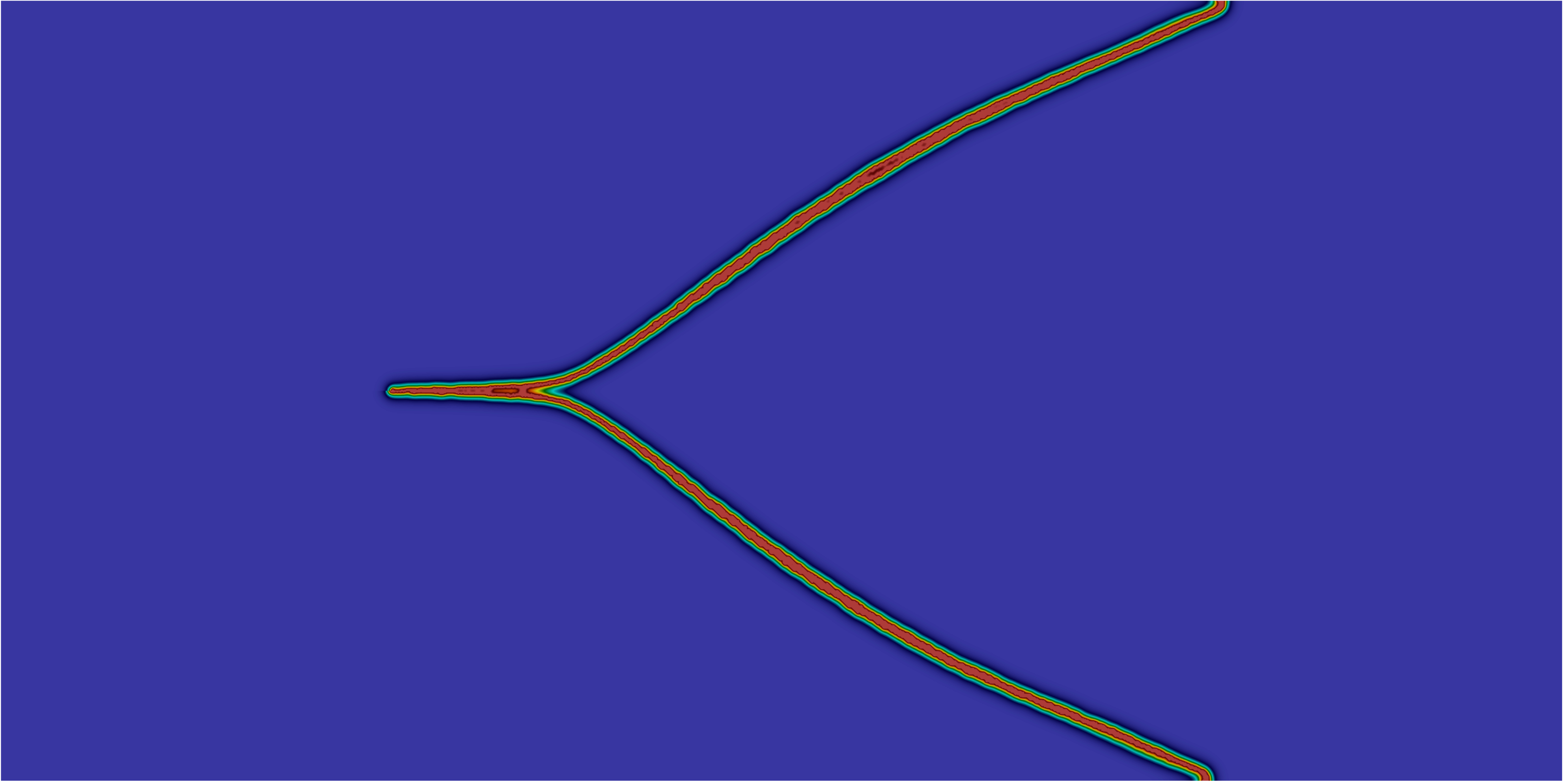}}
\caption{Structured mesh, $t = 173 \, ms$}
\end{subfigure}
  \caption{Space-and-time adaptive simulation of branching fracture
    (cubic degradation function): Phase-field evolution}
\label{Fig:TimeMeshirregPhaseF}
\end{figure}

\subsection{Dissipation analysis and initial mesh sensitivity}

We now focus on the bias induced by the initial mesh. We analyze this influence by comparing structured versus unstructured meshes. We consider a cubic degradation function in this case, inducing the fracture with a traction force $\boldsymbol{\sigma}\left(t\right) = 8 \ kN$. We compare the dissipated energy for both mesh types, where we compute this energy following~\cite{ SARGADO2018458}, that is, 
\begin{equation}
  \mathcal{D} = \int_{\mathcal{B}_{0}} \frac{G_c}{2} \left(
    \frac{1}{\ell} \left(1
      - \varphi \right)^2 + \ell \| \nabla \varphi \|^2 \right) \ d\mathcal{B}_{0}.
  \label{eq:dissipation}
\end{equation}

Figure~\ref{Fig:irregularinitialmesh}~(a) shows the evolution for an unstructured mesh, starting with $2,464$ elements. By construction, this mesh contains highly irregular elements near the notch tip; see, for example, Figure~\ref{Fig:irregularinitialmesh}~(b). We use the same numerical and material parameters as above with a cubic degradation function instead of a quadratic one as in the previous cases.

Figure~\ref{Fig:irregularinitialmesh}~(a) shows the temporal evolution of the element number and the mesh size for both types of meshes. In the case of the structured mesh, the refinement starts earlier than in the unstructured one since the crack tip's resolution needs to improve to capture the crack onset accurately. The final element numbers are $195,273$ for the unstructured mesh and $166,529$ for the structured one. We force the refinement procedure to have at least five elements to reproduce the phase-field interface in both cases.

Furthermore, Figure~\ref{Fig:irregularinitialmesh}~(b)  shows the free fracture energy evolution for both meshes. Although the structured mesh starts propagating the fracture after the unstructured one, the dissipation velocity is higher for the structured mesh. Our method shows is irreversible, that is,
\begin{equation}
\mathcal{D}\left(t_{n+1}\right) \geq 0,
\end{equation}
implying that the system has a free fracture energy at time $t_{n+1}$  that is lower or equal than one at the previous time step.

Figure~\ref{Fig:TimeMeshirreg} shows the snapshot sequence of both initial meshes, while Figure~\ref{Fig:TimeMeshirregPhaseF} shows the phase-field evolution for both cases. The final mesh and phase-field configuration for both cases are almost identical, demonstrating the method's robustness even when starting from highly irregular meshes.


\section{Conclusions} \label{sec:conclusion}

We present a space-and-time adaptive method for dynamic fracture problems based on Eulerian-Lagrangian formulations. First, we describe a thermodynamically consistent fracture model based on phase-field theory, which allows us to formulate from an energetic point of view. Then, we describe a staggered solution scheme that uses second-order generalized-$\alpha$ time marching methods for first- and second-order time derivatives. Furthermore, we detail a time adaptive method that estimates the temporal error using backward difference formulas from previously computed time steps and the generalized-$alpha$ methods' updates. This strategy results in a simple equation that calculates the truncation error based on previous solutions of the equilibrium equations; this truncation error via a weighted truncation estimate allows us to design an error-based time adaptive process. We combine this time adaptive method with an adaptive mesh method based on a residual minimization based on the phase-field equation to deliver a fully automatic space-and-time adaptive strategy for dynamic fracture simulation. We use a bubble-enriched finite element space to estimate the residual error in a proper norm. The general algorithm we propose solves all equations in the proposed residual-minimization scheme at each time step. We detail important algebraic aspects and the refinement criteria we consider for this class of dynamic fracture branching problems.  

We use three challenging dynamic fracture propagation problems to demonstrate the efficiency and robustness of our method. First, we study the robustness of our temporal error estimation for time adaptivity in a notched plate subject traction that induces a dynamic cracking process. We use a regular mesh with $262,144$ elements and a mesh size of $\ell =5 \, mm$. In particular, we compare the two time-adaptive strategies, one driven by the iteration count, against our error-based approach. Our tests show that the iteration-count scheme leads to asymmetrical solutions in symmetric problems. Our error-based time-adaptive strategy is robust and delivers symmetric solutions on the same mesh. The second example solves the same problem and studies the influence of the adaptive mesh strategy on the simulation solutions for this case. In this case, we start from a coarse mesh and allow the mesh adaptivity to track the crack path during the fracture's dynamic evolution. Our space-and-time adaptive scheme requires a We obtain a final mesh with $115,653$ elements with a mesh size $\ell =0.95 \, mm$. This final mesh uses less than half the elements the regular mesh requires. The method significantly reduces overall computational cost while delivering a better resolution of the crack path and the crack tip dynamics.  The last example analyzes the energy dissipation of the adaptive strategy, showing that the proposed algorithm respects the problem's irreversible nature.
In particular, we consider a cubic degradation function and study the dynamic crack evolution that structure and unstructured meshes deliver. We use meshes with a similar number of degrees of freedom. We build the structured mesh to be regular with smooth element size transitions. In contrast, the unstructured mesh is highly irregular, the product of automatic mesh generation. Nevertheless, the mesh bias introduced by the initial mesh is negligible; our formulation deals even with initial meshes that contain almost flat elements. In conclusion, we show that our method is robust, accurate, and computationally efficient for fracture problems involving inertial effects in the context of Euler-Lagrangian formulations.





\bibliographystyle{unsrt}
\bibliography{bib,selfbib}

\end{document}